%% file: main.tex
\documentclass[acmsmall,screen]{acmart}

\input{styles}

\input{macros}
\setcopyright{rightsretained}
\acmPrice{}
\acmDOI{10.1145/3591274}
\acmYear{2023}
\copyrightyear{2023}
\acmSubmissionID{pldi23main-p331-p}
\acmJournal{PACMPL}
\acmVolume{7}
\acmNumber{PLDI}
\acmArticle{160}
\acmMonth{6}
\received{2022-11-10}
\received[accepted]{2023-03-31}
\begin{document}

\title{Search-Based Regular Expression Inference on a GPU}

\author{Mojtaba Valizadeh} \email{Valizadeh.Mojtaba@gmail.com}
\orcid{https://orcid.org/0000-0003-1582-3213} \affiliation{
  \institution{University of Sussex} \city{Brighton} \country{UK} }

\author{Martin Berger} \email{contact@martinfriedrichberger.net}
\orcid{https://orcid.org/0000-0003-3239-5812} \affiliation{
  \institution{Montanarius Ltd} \city{London} \country{UK} }
\affiliation{ \institution{Huawei R\&D UK Ltd} \city{London}
  \country{UK} } \affiliation{ \institution{University of Sussex}
  \city{Brighton} \country{UK} }

\renewcommand{\shortauthors}{Mojtaba Valizadeh \& Martin Berger} 

\input{abstract}
\input{ccs} 

\maketitle
\input{introduction}

\input{mathematicalPreliminaries}

\input{algorithm}

\input{measurements}

\input{conclusion}

\input{acknowledgement}
\input{data-availability}

\bibliographystyle{ACM-Reference-Format}
\bibliography{bib} 

\end{document}

%% file: styles.tex
\usepackage{dashrule}
\usepackage{amsmath, amsfonts, ,graphicx}
\usepackage{mathrsfs,url}
\usepackage{lscape}
\usepackage{multicol}
\usepackage{algorithm}
\usepackage[noend]{algpseudocode}

\usepackage{xspace}

%% file: macros.tex
\newcommand{\ANONYMISED}[1]{\textbf{ANONYMISED}}

\newcommand{\LEQ}{\sqsubseteq}
\newcommand{\LENEQ}{\sqsubset}

\newcommand{\IC}[1]{\mathsf{ic}(#1)}
\newcommand{\NI}{\noindent}
\newcommand{\OL}[1]{\overline{#1}}

\newcommand{\LANG}[1]{\mathsf{Lang}(#1)}

\newcommand{\STRING}[1]{\langle #1 \rangle}
\newcommand{\BOOL}{\mathbb{B}}

\newcommand{\CHARACTERISTIC}[2]{\mathbf{1}^{#1}_{#2}}
\newcommand{\SUPPORT}[1]{\mathsf{supp}(#1)}

\newcommand{\FPS}[2]{#1\STRING{#2}}
\newcommand{\POLYNOMIALS}[2]{#1\STRING{\!\STRING{#2}\!}}
\newcommand{\INFIXFPS}[2]{\FPS{#1}{#2}^{ic}}

\newcommand{\EXTRACT}[2]{\mathsf{extract}(#1, #2)}

\newcommand{\DEF}{Def.\ } 
\newcommand{\WRT}{w.r.t.\ } 
\newcommand{\IE}{i.e.,\ } 
\newcommand{\EG}{e.g.,\ } 
\newcommand{\EEG}{E.g.,\ } 
\newcommand{\ALG}{Alg.\ } 
\newcommand{\SEC}{Sec.\ } 

\newcommand{\AREPO}{https://github.com/martinberger/regular-expression-learning/blob/master/anonymous-repo-for-all-artefacts}
\newcommand{\ALPHAREGEXREPO}{https://github.com/kupl/AlphaRegexPublic}
\newcommand{\ALPHAREGEX}{\textsc{AlphaRegex}\xspace}
\newcommand{\AR}{$\alpha$\textsc{R}\xspace}
\newcommand{\ALGO}{\textsc{Paresy}\xspace}
\newcommand{\ALGOSHORT}{\ALGO}

\newcommand{\TYPEFONT}[1]{\mathsf{#1}}
\newcommand{\RE}[1]{\mathsf{RE}(#1)}
\newcommand{\REGLANG}[1]{\mathsf{Reg}(#1)}

\newcommand{\NAT}{\TYPEFONT{Nat}}

\newenvironment{FIGURE}{\begin{figure}}{\end{figure}}
\newenvironment{TABLE}{\begin{table}}{\end{table}}
\newcommand{\FS}{\rightarrow}
\newcommand{\EMPH}[1]{\emph{#1}}
\newenvironment{GRAMMAR}{\[\begin{array}{lcl}}{\end{array}\]}

\newcommand{\VERTICAL}{\  \mid\hspace{-3.0pt}\mid \ }

\newcommand{\CARD}[1]{\# #1}
\usepackage{ifthen}
\newboolean{showcomments}
\setboolean{showcomments}{true}
\ifthenelse{\boolean{showcomments}}
  {\newcommand{\mynote}[2]{
    \fbox{\bfseries\sffamily\scriptsize#1}
    {\small$\blacktriangleright$\textsf{\emph{#2}}$\blacktriangleleft$}
   }
  }
  {\newcommand{\mynote}[2]{}
  }

\newcommand{\POWERSET}[1]{\mathfrak{P}(#1)}

\newcommand{\COST}[1]{\mathsf{cost}(#1)}
\newcommand{\CONG}{\simeq}
\newcommand{\LENGTH}[1]{|\!|#1|\!|}
\newcommand{\NOVSPACEPARAGRAPH}[1]{\NI\subsubsection*{#1}}
\newcommand{\PARAGRAPH}[1]{\vspace{2mm}\NOVSPACEPARAGRAPH{#1}}

%% file: abstract.tex
\begin{abstract}
  Regular expression inference (REI) is a supervised machine learning
  and program synthesis problem that takes a cost metric for regular
  expressions, and positive and negative examples of strings as input.
  It outputs a regular expression that is \EMPH{precise} (\IE accepts
  all positive and rejects all negative examples), and \EMPH{minimal}
  \WRT to the cost metric. We present a novel algorithm for REI over arbitrary alphabets that is
  enumerative and trades off time for space.  Our main algorithmic
  idea is to implement the search space of regular expressions
  succinctly as a contiguous matrix of bitvectors.  Collectively, the
  bitvectors represent, as characteristic sequences, all sub-languages
  of the infix-closure of the union of positive and negative examples.
  Mathematically, this is a semiring of (a variant of) formal power
  series. Infix-closure enables bottom-up compositional construction
  of larger from smaller regular expressions using the operations of
  our semiring. This minimises data movement and data-dependent
  branching, hence maximises data-parallelism. In addition, the
  infix-closure remains unchanged during the search, hence search can
  be staged: first pre-compute various expensive operations, and then
  run the compute intensive search process.  We provide two C++
  implementations, one for general purpose CPUs and one for Nvidia
  GPUs (using CUDA). We benchmark both on Google Colab Pro: the GPU
  implementation is on average over 1000x faster than the CPU
  implementation on the hardest benchmarks.
\end{abstract}

%% file: ccs.tex
\ccsdesc[500]{Theory of computation~Inductive inference}
\ccsdesc[500]{Computer systems organization~Single instruction, multiple data}
\ccsdesc[500]{Theory of computation~Regular languages}
\ccsdesc[500]{Theory of computation~Massively parallel algorithms}

\keywords{Grammar inference, regular expression inference, program synthesis, machine learning, GPU.}

%% file: introduction.tex
\section{Introduction}\label{introduction}

\NI This paper answers the following quantitative research question in the affirmative:

\begin{quote}
  \EMPH{Can well-known machine learning approaches,
    other than neural networks, benefit from GPU acceleration, in the
    sense of running at least 2 orders of magnitude faster than
    comparable CPU implementations?}
\end{quote}

\NI What is the technical essence that allowed GPU acceleration of
artificial neural networks (ANNs)?  Simplifying a great deal, computer
graphics algorithms often have the following characteristics:
\begin{itemize}

\item Highly parallel.
\item Predictable data movement due to high spatial and temporal data locality.
\item Little to no conditional execution based on non-local data.

\end{itemize}
Specialising processor architecture to such algorithms avoids the
overhead of general purpose CPUs (\EG branch prediction, complex cache
hierarchies and out-of-order execution)
\cite{DallyWJ:domspeha,Hennessy17Computer} and leads to
GPUs. Generalised matrix multiplication algorithms (GeMM) have similar
advantages, which explains the popularity of GPUs in scientific
computing.  Training and inference of ANNs can also be reduced to
GeMM, which we believe is why they can be accelerated on GPUs.  As
far as we are aware, few other well-known machine learning
(ML) techniques are currently known to be accelerable on GPUs. Can we
change this?  In answer to this question, this paper considers precise
and minimal regular expression inference (REI) from positive and
negative examples of strings. REI, the most well-studied form of
grammar inference \cite{Wikipedia:graind}, is one of the oldest
approaches to ML \cite{GoldEM:lanideitl,AngluinD:learegsfqac}.  
To the best of our knowledge, precise and minimal REI is an
open problem for ANNs.

Regular expressions are a constrained mechanism for succinct, finite
specification of finite and infinite languages.  While all finite
languages are definable, regular expressions can only specify simple
infinite languages.  Regular expressions are one of the most widely
used and well-known formalisms in computer science, and we assume that
the reader is familiar with them. Here is an example of a regular
expression
\[
   10(0 + 1)^*
\]
which specifies the language of all strings with characters from $\{0,
1\}$ that start with 10. We assume that regular expressions $r$ have
a \EMPH{cost}, written $\COST{r}$. The exact nature of $\COST{\cdot}$
will play a key role later, for now a naive understanding of cost, for
example the length of $r$ as a string, is sufficient.  It is often
bothersome to write down regular expressions explicitly, so
instead we'd like to program them ``by example'': we give a set of
positive and negative examples of strings, and get a suitable regular
expression that accepts all positive examples and rejects all negative
examples.  For example
\begin{align}\label{running_example_132}
\begin{array}{lcl}
\text{Positive} &:&
10, \ 
101, \ 
100, \ 
1010, \ 
1011, \ 
1000, \ 
1001 \\
\text{Negative} &:& 
\epsilon, \
0, \
1, \
00, \
11, \
010
\end{array}
\end{align}
should, ideally, lead to the target expression $10(0 + 1)^*$.
Clearly this is a special case of supervised learning, and, like all
supervised learning, is subject to subtle problems, including that
there isn't a unique way of generalising from a finite set of
examples. The regular expression
\[
   10(0^* + 1^*)^* + 1000
\]
and infinitely many others, correctly accept all positive examples and
rejects all negative examples from (\ref{running_example_132}). We want
a canonical ``natural'' regular expression to be inferred from the
example. But what does it mean to be canonical and natural?  One might
argue that the regular expression
\begin{align}\label{running_example_132_overfit}
   10 + 101 + 100 + 1010 + 1011 + 1000 + 1001
\end{align}
is both, natural and canonical, and indeed minimal in some sense (it
accepts exactly the positive examples and rejects every other string).  At the same
time it is unlikely to be what the author of the examples had in mind.
Instead, it, in the language of modern ML, \EMPH{overfits} on the
examples.  The ever-present option to overfit shows that REI is trivial without
additional constraints such as minimality.  With regular expressions,
overfitting is easy to avoid, an insight we adapt from
\cite{Feser:JK:syndatstrtfioe}, since we can request a  minimum
cost regular expression that meets all examples. We reject
 (\ref{running_example_132_overfit}) as it is much bigger
than $10(0 + 1)^*$. Such minimisation can be seen as a form of
regularisation, but there are multiple ways of measuring a regular
expression's cost, so minimality is always relative to a chosen cost
measurement. We come back to this later.

The core insight that lets us implement fast REI on a GPU is the
following.  First, instead of regular expressions, we search over
regular languages, which are certain functions of type $r : \Sigma^*
\FS \{0, 1\}$, where $\Sigma$ is the ambient alphabet (of arbitrary size), and $r(w) = 1$
iff $w$ is in the language.  Ignoring cost, two regular expressions
are equal with respect to a pair of sets of positive and negative examples, $P$ and $N$, if
their respective languages relate to the members of $P \cup N$ in
the same way.  Hence, during search, we can restrict our attention
to functions $(P \cup N) \FS \{0, 1\}$, but, in order to build up
these functions in a compositional way that preserves cost-minimality,
we instead search over a small generalisation:
\[
   \IC{P \cup N} \FS \{0, 1\}
\]
Here $\IC{S}$ is the infix-closure of a set ($w$ is an infix, aka substring, of 
string $\sigma$, if $\sigma$ is of the form $\sigma_1 w \sigma_2$ for
some strings $\sigma_i$). Since computer memory is totally ordered, we
can implement each such function as a bitvector. Since $P$ and $N$ do
not change during each REI run, all bitvectors that arise during
search have the same length, and collectively form a binary matrix in
memory.  This representation allows us to implement REI mostly using
matrix operations with little data-dependent branching, predictable
data movement and enables a great deal of parallelism.

\PARAGRAPH{Contributions}
In summary, our contributions are as follows:

\begin{itemize}

\item A data parallel algorithm for precise and minimal REI from
  positive and negative examples that is based around succinct
  representation of data structures and trades off memory for speed,
  to minimise data movement and data-dependent branching. Most
  algorithmic choices work for general grammar inference, and are not
  specific to regular expressions.

\item Implementations of the algorithm on CPUs and GPUs.

\item A parameterised benchmark suite with examples that we believe
  are useful for evaluating the performance of REI, beyond the present
  paper.

\item Performance measurements of our implementations showing that
  both are faster than existing algorithms and that our GPU version
  is orders of magnitude faster than the CPU version. All measurements
  are available from \cite{ArtifactGithub}.

\item A mathematical foundation for our algorithm based on the
  well-established theory of formal power series.

\end{itemize}

%% file: mathematicalPreliminaries.tex
\section{Mathematical preliminaries}\label{formal}\label{preliminaries}



In order to establish terminology, we begin with a condensed review of
standard mathematical concepts used later.

\subsection{Background}
\begin{definition}
We assume that $\NAT = \{0, 1, 2, ...\}$.  By $\BOOL$ we denote the set $\{0, 1\}$ of Booleans. We use $0$ for falsity and  $1$ for
truth.  We write $\POWERSET{A}$ for the \EMPH{powerset} of $A$.  We
write $\CARD{S}$ for the cardinality of the set $S$. The
\EMPH{characteristic function} of a set $S \in \POWERSET{A}$ is the
map $\CHARACTERISTIC{A}{S} : A \FS \BOOL$ which maps $a \in A$ to 1
iff $a \in S$, and otherwise to $0$. If $A$ can be disambiguated from
the context, we write $\CHARACTERISTIC{}{S}$ for
$\CHARACTERISTIC{A}{S}$.
  By the \EMPH{average} of a list $x_1, ..., x_n$ of numbers we mean
  their arithmetic mean, \IE $\frac{1}{n}(x_1 + \cdots + x_n)$.
\end{definition}




\begin{definition}\label{def_monoid}
  A \EMPH{monoid} is a tuple $(M, \circ, \epsilon)$ such that $M$ is a
  set, $\circ : M^2 \FS M$ is an associative function, and $\epsilon
  \in M$ such that: for all $m \circ \epsilon = m = \epsilon \circ m$.
  A set $S \subseteq M$ is \EMPH{infix-closed} if whenever $a \circ b
  \circ c \in M$ then also $b \in M$. By $\IC{S}$, the \EMPH{infix-closure} of
  a set $S$, we mean the smallest infix-closed superset of $S$.
A \EMPH{semiring} is a tuple $(S, +, \circ, 0, 1)$ such that: $(S, +, 0)$ is
a commutative monoid, $(S, \circ, 1)$ is a monoid, the distributive
laws hold, \IE $a\cdot (b+c)=(a\cdot b)+(a\cdot c)$ and $(a+b)\cdot
c=(a\cdot c)+(b\cdot c)$,  and, finally, $0\cdot a=0=a\cdot
0$.  We call $\circ$ and $+$ the \EMPH{product} and \EMPH{sum} of the
semiring.
\end{definition}


\begin{definition}
  An \EMPH{alphabet} is a finite set $\Sigma$, and we often speak of
  \EMPH{characters} when we mean the elements of $\Sigma$. A \EMPH{string
    of length $n \in \NAT$} over $\Sigma$ is a map $\sigma : \{0, 1, ..., n-1\} \FS A$.
  We write $\LENGTH{\sigma}$ for $n$. A \EMPH{string over $\Sigma$} is a
  string of length $n$ for some $n$. We often omit stating the
  alphabet where this is clear from the context.  We often write
  $\sigma_i$ instead of $\sigma(i)$. We write $\sigma \cdot \gamma$,
  or just $\sigma\gamma$ for the concatenation of two strings $\sigma$
  and $\gamma$. We write $\epsilon$ for the unique string of length
  $0$.  We write $\Sigma^*$ for the set of all strings over $\Sigma$, and $\Sigma^n$
  the restriction of $\Sigma^*$ to strings of length $n$. We write $\Sigma^{\leq
    n}$ for the set of strings of length not exceeding $n$.
\end{definition}

\begin{definition}\label{def_language}
  A \EMPH{language} over the alphabet $\Sigma$ is a set of strings
  over $\Sigma$.  We  have the
  following well-known algebraic operations over $\Sigma^*$: constants
  $\emptyset$ and $\epsilon$ (the language containing only the empty string), negation (aka
  complement), $\OL{L}$ for $\Sigma^* \setminus L$, union (aka
  disjunction), $L \cup L'$, concatenation, $L_1 \cdot L_2$, or just
  $L_1L_2$, for $ \{ \sigma \gamma \in \Sigma^*\ |\ \sigma\in L_1,
  \gamma \in L_2 \} $, intersection or conjunction, $L_1 \cap L_2$.
  Finally there is the Kleene-star, where for all $n$: $ L^0 =
  \{\epsilon\} $, $L^{n+1} = L^n \cdot L$ and then $ L^* = \bigcup_{n
    \geq 0}L^n $. Those operations form various algebras, for example
  $(\POWERSET{\Sigma^*}, \cup, \cdot, \emptyset, \epsilon)$ is a
  semiring.

\end{definition}

\begin{definition}
  A \EMPH{partial order} is a tuple $(P, \LEQ)$ where $P$ is a set and
  $\LEQ$, a subset of $P^2$ is a binary, reflexive, anti-symmetric and
  transitive relation on $P$.  We write $p \LENEQ q$ to signify that
  $p \LEQ q$ and $p \neq q$.  We say the order is \EMPH{total} if
  always $p \LENEQ q$, or $q \LENEQ p$ or $p = q$.  If $Q \subseteq P$
  then $\LEQ$ also orders $Q$ by \EMPH{restriction}.  If a set
  $\Sigma$ is ordered by $\LEQ$, then this order can be lifted to
  $\Sigma^*$ using the \EMPH{shortlex ordering}: $ \sigma \LEQ
  \sigma' $ iff either $\LENGTH{\sigma} < \LENGTH{\sigma'}$, or for
  some $i$ we have: $\sigma_i \LEQ \sigma'_i$ and, at the same time,
  for all $j < i: \sigma_j = \sigma'_j$.  Note that, by restriction,
  this orders every subset of $\Sigma^*$.
\end{definition}
Note that anything stored in a computer's memory is always
totally ordered, since the addresses of memory cells are
integers. 

\begin{definition}\label{def_regular_language}
  The \EMPH{regular languages} over an alphabet $\Sigma$, denoted
  $\REGLANG{\Sigma}$, are inductively given by the following
  constraints: the empty set is regular; for each $a \in \Sigma$, the
  language \{a\} is regular; if $L_1$ and $L_2$ are regular then so
  are $L_1 \cup L_2$ and $L_1 \cdot L_2$.; if $L$ is regular then
  $L^*$ is regular.
\end{definition}

\begin{definition}\label{def_regular_expressions}
  The \EMPH{regular expressions} over $\Sigma$, short
  $\RE{\Sigma}$, are given by the following grammar:
  \begin{GRAMMAR}
    r
    &::=&
    \emptyset
    \VERTICAL
    \epsilon
    \VERTICAL
    a
    \VERTICAL
    r \cdot r
    \VERTICAL
    r + r
    \VERTICAL
    r^*
  \end{GRAMMAR}
  Here $a$ ranges over $\Sigma$.  We call the $*, \cdot, +, ...,
  \emptyset$ the \EMPH{regular constructors (over $\Sigma$)} of regular
  expressions, where each regular constructor has the obvious \EMPH{arity},
  \EG $*$ has arity 1, while $+$ has arity 2.  We use various
  abbreviations, including $rr'$ for concatenation $r \cdot r'$,
  and $\Sigma^*$ for $(a_1 + \dots + a_k)^*$ assuming that $\Sigma
  = \{a_1, ..., a_k\}$.  
\end{definition}


\begin{definition}
  With each $r \in \RE{\Sigma}$ we associate the
  \EMPH{denotation} of $r$, aka the \EMPH{language of $r$}, abbreviated
  $\LANG{r}$, which is defined by the following clauses:
 $\LANG{\emptyset} = \emptyset$,
 $\LANG{\epsilon} = \{\epsilon\}$,
 $\LANG{a} = \{a\}$,
 $\LANG{r \cdot r'} = \LANG{r} \cdot \LANG{r'}$,
 $\LANG{r + r'} = \LANG{r} \cup \LANG{r'}$,
 $\LANG{r^*} = \LANG{r}^*$.
  This induces an equality on regular expressions: $r$
  is equivalent to $r'$ iff $\LANG{r} = \LANG{r'}$, for example $r+r
  \CONG r$, or $r^{**} \CONG r^*$.  Note that each equivalence class
  has an infinite number of inhabitants.
  We write $r?$ for the regular expression with the same language as $\epsilon +  r$.
\end{definition}



\subsection{Key Structure: Formal Power Series}
 Formal power series (FPS) generalise characteristic functions
 $\CHARACTERISTIC{\Sigma}{L} : \Sigma^{*} \FS \BOOL$ of formal
 languages $L$ to functions $\Sigma^{*} \FS S$ where $S$ is a
 semiring.  This is interesting for us, because well-behaved sets of
 such functions form semirings themselves, so the semiring structure
 on $S$ can be lifted to FPS, see
 \cite{GolanJS:semandta,DrosteM:semforps,SalomaaA:auttheaofps,BerstelJ:ratseratl}.
 \ALGO's core data structure is a generalisation of FPS.

\begin{definition}\label{definition_formal_power_series}
  Let $\Sigma$ be an alphabet and $S$ a semiring. A formal power series is a map
  \[
  r : \Sigma^* \FS S
  \]
  The \EMPH{support} of $r$, written $\SUPPORT{r}$ is the set $
  \SUPPORT{r} = \{m \in \Sigma^*\ |\ r(m) \neq 0\} $ A
  \EMPH{polynomial} is a formal power series with finite support. We
  now make the following definitions. $\FPS{\Sigma^*}{S} $ denotes the
  set of all formal power series $r : \Sigma^* \FS S$.
  $\POLYNOMIALS{\Sigma^*}{S}$ is the subset of $\FPS{\Sigma^*}{S}$,
  but restricted to finite support.  We define the following
  operations on $\FPS{\Sigma^*}{S}$, and, by restriction, on
  $\POLYNOMIALS{\Sigma^*}{S}$, for $\sigma \in \Sigma^*$.
  \begin{itemize}
  \begin{minipage}[t]{0.3\linewidth}
  \item $0(\sigma) = 0$
  \item $1(\sigma) =
    \begin{cases}
      1 & \sigma = \epsilon \\
      0 & \text{else}
    \end{cases}$
  \end{minipage}
  \begin{minipage}[t]{0.7\linewidth}
  \item $(r + s)(\sigma) = r(\sigma) + s(\sigma)$
  \item $(r \cdot s)(\sigma) = \oplus \{ r(\sigma_1) \cdot s(\sigma_2) \ |\ \mathbf{ \sigma_i \in \Sigma^*},  \sigma_1 \cdot \sigma_2 = \sigma \}$
  \end{minipage}
    \end{itemize}
\end{definition}
\NI Here $\oplus$ is semiring addition lifted to finite sets.
We note the formal similarity of $r \cdot s$ with convolutions.
 Under mild restrictions,
which hold in our use-cases, we can define Kleene-star on
$\FPS{\Sigma^*}{S}$ and $\POLYNOMIALS{\Sigma^*}{S}$,
details omitted for brevity.

While \ALGO can be presented as code, we emphasise the connection with
well-known mathematical structures, because they give us a shared
language: semirings and polynomials are widely known and they are a
minimal 'API' for grammar synthesis; ANNs are generalised matrix
operations that also form semirings; and, finally, it puts in context
what is REI specific (\EG infix-closure), and what is not (almost
everything else works for arbitrary formal languages).

%% file: algorithm.tex
\section{The \ALGO algorithm}\label{algorithm}

This section presents \ALGO, our \EMPH{parallel regular expression
  synthesiser}.  To save space we omit details, in particular
about low-level optimisations. They can be found in  the C++
implementations.

\PARAGRAPH{Specifications and cost homomorphisms}
The input to the algorithm is a cost homomorphism, defined below, and a set of
positive and negative examples (over an arbitrary alphabet). 

\begin{definition}\label{def_specification}
  Given an arbitrary alphabet $\Sigma$, a \EMPH{specification (over
    $\Sigma$)} is a pair $(P, N)$ where both $P$ and $N$ are finite
  subsets of $\Sigma^*$. We say a language $L \subseteq \Sigma^*$
  \EMPH{satisfies} $(P, N)$, written $ L \models (P, N) $, provided $P
  \subseteq L$ and $N \cap L = \emptyset$. We say a regular expression
  $r$ \EMPH{satisfies} $(P, N)$, written $r \models (P, N)$, if
  $\LANG{r} \models (P, N)$.
\end{definition}

\begin{definition}
  A \EMPH{cost function} is a map $ \COST{\cdot} : \RE{\Sigma} \FS
  \NAT$. It is a \EMPH{cost homomorphism} if there are
  integer constants, $c_1, ..., c_5 >
  0$, such that $\COST{\emptyset} = \COST{\epsilon} = \COST{a} = c_1$ for all  $a \in \Sigma$,
  $\COST{r?} = \COST{r} + c_2$, $\COST{r^*} = \COST{r} + c_3$, $\COST{r \cdot r'} = \COST{r} +
  \COST{r'} + c_4$ and $\COST{r + r'} = \COST{r} + \COST{r'} + c_5$.
  We call each $c_i$ the \EMPH{cost} of the corresponding
  regular constructor.
\end{definition}
We write, \EG $\COST{*}$ for $c_3$ and likewise for the other costs.
From now on, whenever we present a 5-tuple of numbers,
\EG $(c_1, c_2, c_3, c_4, c_5)$, this is short for the cost
homomorphism $(\COST{a}, \COST{?}, \COST{*},
\COST{\cdot}, \COST{+})$ in this exact order, \EG in $(5, 2, 7, 2,
19)$, the cost of the Kleene-star is 7. Note that we allow $\COST{r?}
\neq \COST{\epsilon} + \COST{r}$, this is for convenient comparison
with related work later. As long as all costs remain strictly positive,
more complex cost homomorphisms can easily be accommodated, \EG
different costs for different alphabet characters.

\PARAGRAPH{Core intuitions}
REI is a search problem over $\RE{\Sigma}$, the syntax of regular
expressions. We have to decide how to represent the search space
$\RE{\Sigma}$ in an implementation.  The natural answer, using
$\RE{\Sigma}$ itself, is wasteful for several reasons.

\begin{itemize}
\item Redundancy: each regular language is denoted by infinitely many
  regular expressions. For example $00 + 1$ and $1 + 00$ denote the
  same language.

\item Not succinct: each regular expression is
  a tree (\IE requiring additional pointers).

\item Slow contains-check: the search will carry out many
  \EMPH{contains-checks}, to determine if a candidate expression
  accepts or rejects a given string.  Depending on implementation
  details, this amounts to expensive 'walking' of the tree
  representing the candidate.

\end{itemize}
In order to avoid those inefficiencies, we 
  represent regular
expressions, simplifying a bit, by their languages,
\IE the search space is (a subset of)
$\LANG{\Sigma}$.  In memory, we could  represent each language $L$
by its  \EMPH{characteristic function}
\[
     \CHARACTERISTIC{}{L} : \Sigma^* \FS \BOOL
\]
which is formal power series in the sense of \DEF
\ref{definition_formal_power_series}.
Mathematically, a function is an unordered set of pairs. Since computer
memory is a totally ordered sequence of bits, we get a total order on
$ \Sigma^*$ (\EG using the shortlex order to lift a chosen
total order on $\Sigma$).  Hence we can represent
$\CHARACTERISTIC{}{L}$ as a list of 0s and 1s in memory. We call this
list \EMPH{characteristic sequence} (CS).  This turns
every language into a bitvector, albeit infinitely long.  Fortunately,
we only need to implement a finite segment of these characteristic
functions: the algorithm returns an $r$ with $r \models (P, N)$.  As
we represent regular expressions by $L$, that means we need
to check $L \models (P, N)$, \IE  only words in 
 $P \cup N$, a finite set. Hence we can represent 
languages as finite functions
\[
    \CHARACTERISTIC{}{L} : (P \cup N) \FS \BOOL
\]
which amounts to a bitvector of length $\CARD{(P \cup N)}$.  In the
rest of this text we will not carefully distinguish between a language
$L$, and its representations as function $\CHARACTERISTIC{}{L} : (P
\cup N) \FS \BOOL$,  bitvector or CS.

We need not just synthesise a regular expression meeting the target
specification, but a minimal one.  In order to do this we lift
the ambient cost function to regular languages.

\begin{definition}
  Given $\COST{\cdot}$, we set $\COST{L}$ to $\COST{r}$
  for a minimal $r$ with $\LANG{r} = L$.
\end{definition}

We now present a key insight that does not seem to appear in the
extensive literature on formal power series and semirings, the
mathematics behind our implementation.

\begin{lemma}\label{lemma_build_up}
  Assume  $\COST{\cdot}$ is cost homomorphism, then for all $L, L_1, L_2\in \REGLANG{\Sigma}$:
 $  \COST{L^*} \leq  \COST{L} + \COST{*}$,
 $  \COST{L?} \leq  \COST{L} + \COST{?}$,
 $  \COST{L_1 \cdot L_2} \leq  \COST{L_1} + \COST{L_2} + \COST{\cdot}$,
 $  \COST{L_1 + L_2} \leq \COST{L_1} + \COST{L_2} + \COST{+}$.
\end{lemma}

\NI This lemma enables compositional, bottom-up construction of
regular languages with increasing cost: in order to construct a
regular language with cost $c$, we choose an outermost regular
constructor, subtract its cost, and recurse. \EEG for $+$
 we split the remaining cost $c - \COST{+}$ into suitable pairs
$c_l$ and $c_r$ and find all languages $L_l$ of cost $c_l$ and $L_r$
of cost $c_r$.  Then $L_l + L_r$ has target cost not exceeding  $c$. Computing the
sum of two languages is just bitwise-or over the corresponding two
bitvectors.  Then we check if $L_l + L_r$ meets the specification. If
not we continue to search, if yes, we reverse engineer a
corresponding regular expression (see below) and return it.
Kleene-star and concatenation are somewhat more complex and discussed
next.

\PARAGRAPH{First space-time trade-off: infix-closure}
There is a problem with using characteristic
sequences $(P \cup N) \FS \BOOL$ for  concatenation and Kleene-star. Recall from \DEF
\ref{definition_formal_power_series} that the product of formal power
series, understood as mappings from $\Sigma^*$ into some semiring, is
given as
\[
(r \cdot s)(\sigma) = \oplus \{ r(\sigma_1) \cdot s(\sigma_2) \ |\  \underline{\sigma_i \in \Sigma^*},  \sigma_1 \cdot \sigma_2 = \sigma \}   
\]
We emphasise the underlined part:
if we define, for $r, s : (P \cup N) \FS \BOOL$
\[
(r \cdot s)(\sigma) = \oplus \{ r(\sigma_1) \cdot s(\sigma_2) \ |\   \sigma_1 \cdot \sigma_2 = \sigma \} 
\]
the question arises: what do $\sigma_1, \sigma_2$ range over? The
answer cannot be: over $P \cup N$. Consider
the specification $(\{01\}, \emptyset)$.  \ALGO works bottom-up,
starting from the lowest cost CS, as we will see below.  The only
way to construct (the CS corresponding to) $0 \cdot 1$ is as
concatenation of (the CS corresponding to) 0 and (the CS
corresponding to) 1, both of which are lower cost than $0 \cdot
1$. But the set of functions on $(P \cup N) \FS \BOOL$ corresponding
to the alphabet characters 0, 1 is empty. One might say: $P \cup N$ is incomplete for bottom-up synthesis!
We are looking for a smallest
but finite superset of $P \cup N$ that is closed under  regular
operations. This motivates the next definition.

\begin{definition}\label{definition_infix_power_series}
  Let $S$ be a semiring, and $I \subseteq \Sigma^*$ be finite and
  infix-closed.  An \EMPH{infix power series} (IPS) is a map
  \[
     r : I \FS S
  \]
  We denote the set of all IPS by $\INFIXFPS{I}{S}$.  We define the
  following operations on $\INFIXFPS{I}{S}$, for $\sigma \in I$.

  \begin{itemize}
  \begin{minipage}[t]{0.3\linewidth}
  \item $0(\sigma) = 0$.
  \item $1(\sigma) =
    \begin{cases}
      1 & \sigma = \epsilon \\
      0 & \text{else}
    \end{cases}$.
  \end{minipage}
  \begin{minipage}[t]{0.7\linewidth}
\item $(r + s)(\sigma) = r(\sigma) + s(\sigma)$.
\item $r^*(\sigma) = \oplus_{n = 0}^{\infty} r^n(\sigma)$.
  \item $(r \cdot s)(\sigma) = \oplus\{r(\sigma_1) \cdot s(\sigma_2) \ |\ \sigma_1, \sigma_2 \in I, \sigma_1 \cdot \sigma_2 = \sigma \}$  
  \end{minipage}
\end{itemize}

Here $r^{0}$ is the characteristic function of the language
$\{\epsilon\}$ and $r^{n+1} = r^{n} \cdot r$.
It is straightforward to show that $\oplus_{n = 0}^{\infty}$
is well-defined because $\IC{P \cup N}$ is finite, see
\cite{DrosteM:semforps} for the more complex general case.  Boolean
operations like negation or conjunction are similarly easy to define,
and omitted for brevity.
\end{definition}

\NI Noting that $(\BOOL, \wedge, \vee, 0, 1)$ forms a
semiring, this gives us \ALGO's search space: $\INFIXFPS{\IC{P \cup
    N}}{\BOOL}$. In other words functions $r : \IC{P \cup N} \FS
\BOOL$, which, with the assumed total order on $\IC{P \cup N}$, give
us bitvectors. The $i$-th element of each bitvector corresponding to $r$ stores the value
$r(w)$, where $w$ is the $i$-th element of $\IC{P \cup N}$.

  \begin{example}\label{example_standard_1}
Consider the specification $P = \{1, 011, 1011, 11011\}$ and $N =
\{\epsilon, 10, 101, 0011\}$. Then $\IC{P \cup N}$ is
\[
 11011, \
 1101, \ 
 110, \ 
 11, \ 
 1011, \ 
 101, \ 
 10, \ 
 1, \ 
 011, \ 
 01, \ 
 0011, \ 
 001, \ 
 00, \ 
 0, \ 
 \epsilon
 \]

  Assume $\IC{P \cup N}$ is ordered as above, and consider the regular
  expression $r = (0?1)^*1$. The intersection of $\LANG{r}$ with
  $\IC{P \cup N}$ is $\{11011, 1011, 011, 11, 1\}$. This can be
  represented as CS, relative to $\IC{P \cup N}$:
\begin{center}  
  \includegraphics[width=0.5\linewidth]{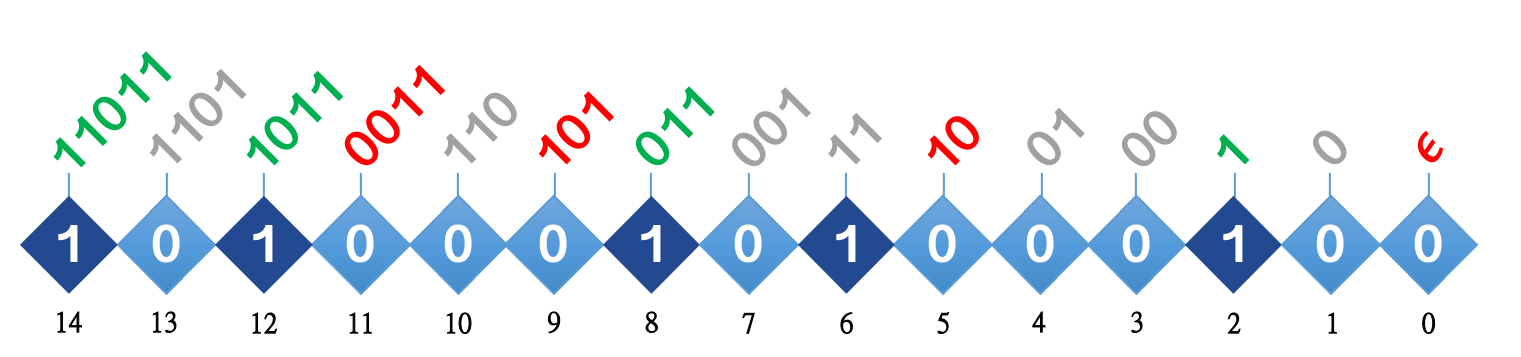}
\end{center}
  \end{example}
  
\NI Here binary strings in green are in $P$, red means $N$, and grey
indicates strings that arise from infix-closure.  The $i$th bit is 1
(dark blue square) exactly when the language contains the $i$th word of
$\IC{P \cup N}$.  Putting multiple CSs contiguously in memory yields a
matrix.
  
\PARAGRAPH{Recovering a regular expression from a regular language}
The algorithm sketched above computes formal languages, but not
regular expressions.  If we build up minimal-cost regular languages
until we find the first $L$ that is compatible with $(P, N)$, we
cannot efficiently, from $L$ alone, produce a minimal regular
expression $r$ such that $\LANG{r} = L$.  We solve this problem by
associating with each bitvector $bv$ enough information to be able to
reverse-engineer a suitable regular expression.  Simplifying at bit,
it means we track the outermost regular constructor used to construct
$bv$ and pointers to the component bitvectors. We store bitvectors produced using the same
outermost regular constructor consecutively, and track, in a
lightweight manner, the information about block beginning / end in a
small external table.  This allows us, recursively, to construct a
regular expression corresponding to the $bv$, on demand.  Details are
fiddly and can be found in the implementation.

\PARAGRAPH{Second space-time trade-off: bitvector length}
How long should bitvectors representing languages be?  If $n =
\CARD{\IC{P \cup N}}$, then $n$  bits suffice, but instead we
choose the smallest power-of-2, not below $n$.  We make this
space-time trade-off because the instruction sets of all modern
processors are designed for operations working on power-of-2 sized
data, typically $8, 16, 32, 64$ bits, sometimes 128 bits. All other
bit-widths must be expressed in terms of those, and hence are much
slower.

\PARAGRAPH{Third space-time trade-off: caching}
If we implement our algorithm naively, lower cost languages will be
recomputed repeatedly when computing higher-cost languages.  We
prevent this with \EMPH{dynamic programming} in the sense of
\cite{FislerK:datcenitc}: we construct all needed languages bottom-up,
from lower to higher cost, and keep the constructed languages in
memory for later re-use in a matrix that we describe below.
This caching is our third space-time
trade-off. It is one of the main reasons for the performance of our
algorithm, but, because the number of regular languages increases
exponentially with increasing cost, makes available memory the
scalability limit. We will see in \SEC \ref{measurement} that on a
modern GPU the algorithm can solve virtually all synthesis tasks in at most a
few seconds, provided they fit in memory.

\PARAGRAPH{Matrix representation: language cache}
During each \ALGO run, $P$ and $N$ are fixed, and so is the size of
$\IC{P \cup N}$.  Hence each potential language, \IE all bitvectors
that arise during the search, have the same length.  We store all 
next to each other in memory, ordered by increasing cost. This amounts
to a matrix called \EMPH{language cache}, the core data structure of
\ALGO.  Ordering the language cache by increasing cost, and noting
that each individual bitvector is itself a one-dimensional matrix,
that means the language cache is a matrix of matrices of matrices,
where the $c$-th entry contains exactly the languages of cost $c$.
The complex, yet regular structure of the language cache allows us to
implement REI mostly using matrix operations with little
data-dependent branching, predictable data movement and enables a
great deal of parallelism. For $(P, N)$ from Example
\ref{example_standard_1}, the language cache could contain  something like
the blue squares in the figure below.

\begin{center}  
	\includegraphics[width=0.65\linewidth]{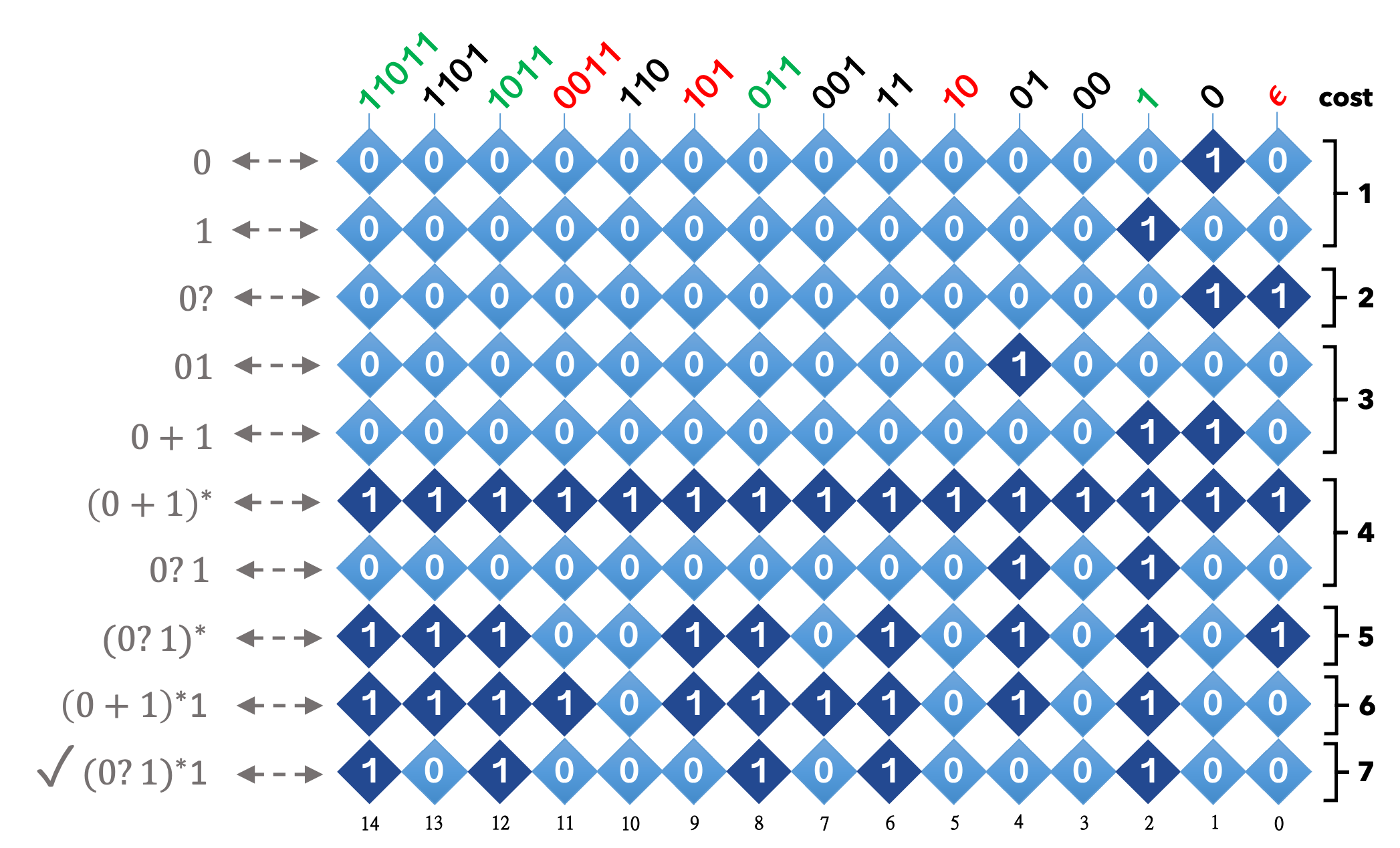}
\end{center}

\NI We annotate every row with a regular expression accepting the
language of the row. Assuming (1, 1, 1, 1, 1) as cost function, the
regular expression is minimal for $(P, N)$. The costs on the right
show the language cache naturally decomposes into clusters of equal
cost.

\PARAGRAPH{Uniqueness checking}
The compositional  construction of languages is not injective: target
languages might be constructed more than once. For example a union of
languages $\{001\} + \{\epsilon\}$ results in the same language as
$\{001, \epsilon\} + \{\epsilon\}$.  In order to avoid the performance
penalty from this duplication we remove them as soon as possible.
Unlike other operations of our algorithm, uniqueness is a global
property: as soon as a new CS is constructed, we compare it to all
previously constructed CS. We add it to the matrix only if it is
genuinely new.  Several things are noteworthy about our approach to
uniqueness checking:
\begin{itemize}

  \item The performance of uniqueness checking is crucial to performance.
  
  \item It works on all formal languages. It is not tied to regular
    languages.

  \item It is subtly different from the \EMPH{pruning} techniques
    proposed in \EG \cite{LeeM:synregefefiaa}, in the sense that they
    prune paths in their search space \EMPH{before} constructing
    regular expressions.  We remove languages \EMPH{after} we
    construct them. Pruning before construction is an interesting
    direction for further work.
    
  \item Computing global properties like uniqueness efficiently on
    GPUs is challenging. Our uniqueness checker is a modified form of
    the \texttt{HashSet} class from  WarpCore
    \cite{JuengerD:warcoralffhtog,JuengerD:warpcoreImplementation}, a
    CUDA library for high-performance hashing of 32 and 64 bit
    integers. We can use WarpCore because we represent as (sequences
    of) unsigned integers (powers-of-2, see above).

  \end{itemize}



    
    
  
    
    








\PARAGRAPH{Staging: guide table}
We have seen how easy it is to compute the union of two formal
languages by bitwise-or.  Fast computation of concatenation or the
Kleene-star is harder because of the convolutional nature of
concatenation, which the Kleene-star iterates. Recall that the product
in $\INFIXFPS{I}{S}$ is defined abstractly as follows:
\[
(r \cdot s)(w) = \oplus\{r(\sigma_1) \cdot s(\sigma_2) \ |\ \sigma_1, \sigma_2 \in I, \sigma_1 \cdot \sigma_2 = w \}
\]
In our case of characteristic sequences $\IC{P \cup N} \FS \BOOL$, the
check $\sigma_1 \cdot \sigma_2 = w$ is somewhat expensive, and, if our
algorithm was implemented naively, would have to be re-run every time
we construct a new characteristic sequence from old using
concatenation or Kleene-star. Fortunately $P, N$ remain constant, and
we pre-compute all ways in which a word $w$ can be split. This amounts
to a function
\[
   gt : \IC{P \cup N} \FS \POWERSET{\IC{P \cup N}^2}
\]
that returns $gt(w) = \{(\sigma_1, \sigma_2)\ |\ \sigma_1 \cdot
\sigma_2 = w\}$. The concrete implementation is the \EMPH{guide table},
an array of arrays of pairs of offsets into the language cache, and we omit
details for brevity.

\PARAGRAPH{OnTheFly mode}
Space-time trade-offs make the algorithm memory intensive. This is the
main scaling limitation. The OnTheFly mode alleviates the problem
somewhat, without compromising on minimality and precision.  The
insight enabling OnTheFly is that, depending on cost function,
computing a regular language of target cost might make reference only
to lower cost CSs that are still cached, \EG if the cost of all
regular constructors is > 55, then the algorithm needs only CSs of
target cost minus 55. That means, even when we have run out of
language cache space, the algorithm can continue for a while, creating
new CSs from old CSs still in the language cache, but these new CSs
are neither cached, not checked for uniqueness.  Avoiding uniqueness
checks makes OnTheFly much faster.  Our algorithm automatically
switches to OnTheFly mode when the language cache is full.
Eventually, OnTheFly mode needs access to CSs that are no longer
cached, then synthesis stops without having found a suitable regular
expression. Our implementation and measurements in \SEC
\ref{measurement} refer to this as out-of-memory error.

\PARAGRAPH{GPU language cache implementation}
The explanations so far used mathematical concepts, because we wanted
to separate low-level implementation details from high-level abstract
structural ideas. However, the block matrix structure of the language
cache, essentially a matrix of matrices of matrices, is itself
interesting, and, especially on GPUs where locality of memory access
matters, important for performance.

The figure below sketches how a new cost-level is built put.  Grey
parts are temporary data, while blue is permanent data. L and R denote
auxiliary data, allowing the conversion of a CS to a corresponding
regular expression. In (a) we see how a new cost level is built up
from data already in the language cache.  The newly constructed CS are
first held in temporary storage. Only CS in temporary memory that
passes the uniqueness check is then copied into the language cache in
(b) filling the new cost level.  The use of intermediate temporary
storage enables parallelism on GPUs.

\begin{center}
\includegraphics[width=0.95\linewidth]{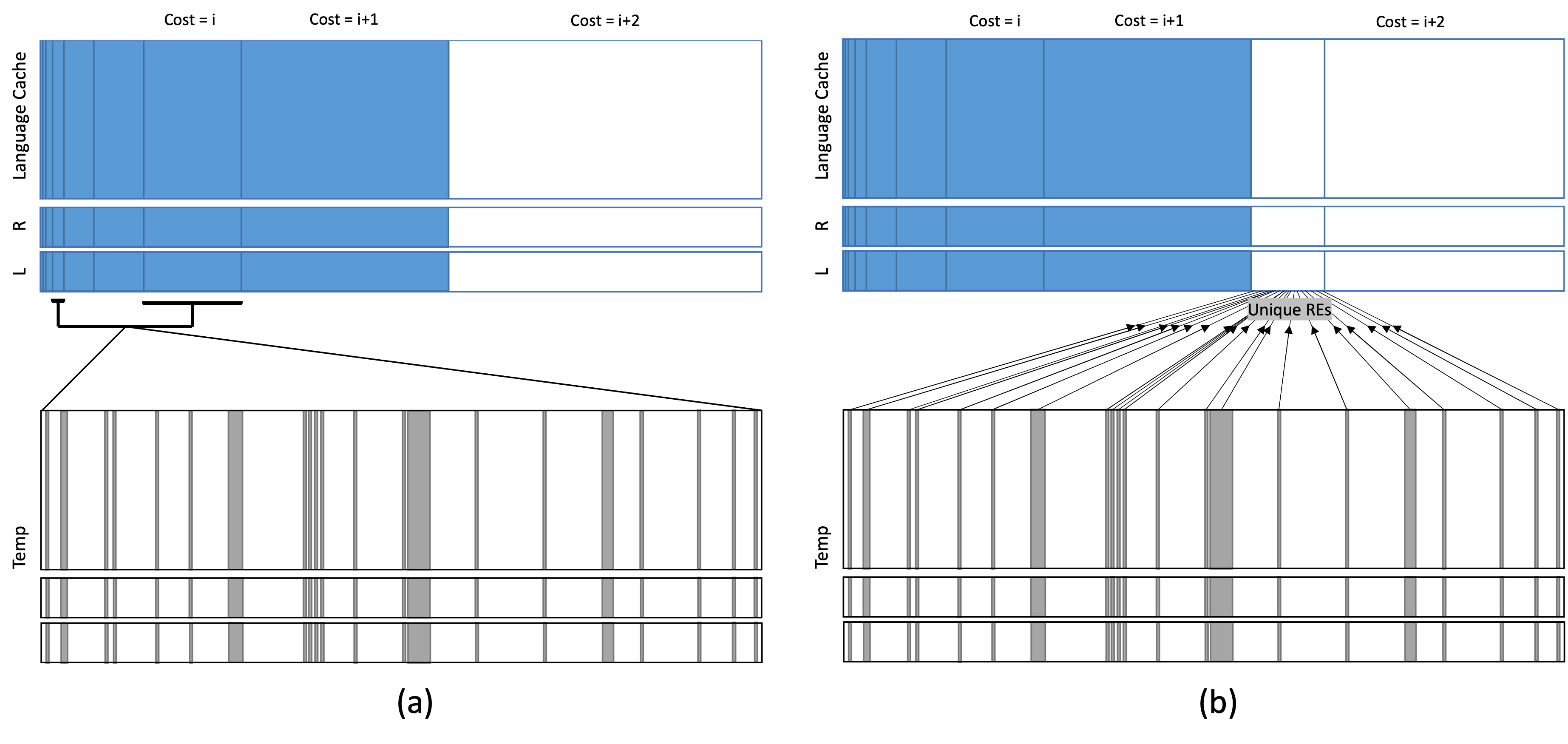}
\end{center}

The overall structure of the algorithm is simple:
\begin{itemize}
  
\item Allocate the language cache as contiguous array of bytes (which
  need not be initialised, as the rest of the algorithm guarantees
  that the first read always happens after the first assignment). The
  internal structure of the language cache as contiguous array emerges
  during search.

\item During searching, the language cache is filled in a single sweep
  from left to right, in a write-once manner.  No element of the
  language cache is ever removed or even changed, once in the language
  cache.

\item The cost of CSs stored in the language cache is never decreasing.

\item A newly created CS is not directly stored in the language
  cache. 
  
\item Search terminates only if the algorithm finds a solution to the
  specification, or OnTheFly has exhausted the language cache's supply
  of CSs.

\end{itemize}

The pseudocode below adds more detail.  For simplicity, pseudocode
models the language cache as an list of fixed size entries, indexed by
cost: accessing the $i$-th language cache entry returns a list of CSs
with cost $c$, we ignore the aforementioned auxiliary information.

{\small \input{code/main} }

\ALG \ref{algo_pseudo_main2} shows the overall structure of \ALGO.
Lines \ref{main:line:init:1} and \ref{main:line:init:2} handle trivial
specifications.  Line \ref{main:line:3} fills the initial language
cache with the initial CSs, corresponding to characters of the
alphabet $\Sigma$, at index $cost(a))$.  The loop then sweeps
over all allowed costs, with increasing cost. For each cost $c$, it
considers all CSs of cost $c$, first with question mark as outermost
constructor, then with  Kleene-star, then with concatenation, and finally 
union.  Each returns either a regular expression that solves
$(P, N)$ and the algorithm terminates (we don't model the trivial
details), or else a list of all \EMPH{new} CSs.  Line
\ref{main:line:9} then concatenates all those new CSs, and makes them
the language cache entry for cost $c$.
As mentioned above, we omit details about having a single contiguous
cache, this is straightforward, but fiddly. (We access the language
cache through a layer of indirection that translates cost into memory
offsets, using a data-structure called \EMPH{startPoints} that is
dynamically updated. This is made easy by the write-once nature of the
language cache.)

\ALG \ref{algo_pseudo_concat} is  pseudocode for constructing CSs
with concatenation as outermost constructor.  It relies on the guide
table, described earlier, which pre-computes all the ways each string
$w$ can be split into strings from $\IC{P \cup N}$.
In code we access the guide table with the index of the target word in $\IC{P
  \cup N}$, rather than the word itself. Line
\ref{concat:line:2} splits the available cost $c$ into all pairs $(L,
R)$ of costs that sum up to $c$. The next two lines then retrieve all
CSs of costs $L$ and $R$ from the language cache. On GPUs this is
done in parallel. For each pair $(lCS, rCS)$ Line
\ref{concat:line:7} loops  over all words in $\IC{P \cup
  N}$. Here $w$ is the index of the word in $\IC{P \cup N}$, which
we assume to be totally ordered.
Line \ref{concat:line:6} initialises
the local variable $newCS$ which is the CS we are constructing. It
is initialised to the empty language.
Line \ref{concat:line:5}
initialises $i$, the 'pointer' into $newCS$ at the position of
$w$. This lets us set the bit at the right place to $newCS$.
In Line
\ref{concat:line:8} we start searching through all guide
table entries for $w$, the (index of the) current word we are
interested in.  Let's say the guide table entry contains $(l, r)$. If
$lCS$ contains the $l$-th word and $rCS$ contains the $r$-th word
(that is checked in Line \ref{concat:line:9}), then that means $w$ is
in the CS under construction, so Line \ref{concat:line:10} sets the
relevant bit $i$.  Once we have finished with the current $w$, we
update our 'pointer' $i$ (by logical left shift) in Line
\ref{concat:line:13}, and process the next word.
Once $newCS$ is constructed, we check if it was already constructed
previously.
If yes, we try with new choices for $lCS$ and $rCS$.
If not, we check if it solves the specification $(P, N)$ and if it
does, the program converts it into a minimal cost RE, and
terminates.
Otherwise we add $newCS$ to the language cache.
For brevity we omit the pseudocode for $buildStar$ which just iterates
concatenation a finite number of times, and $buildUnion$ and
$buildQuestionMark$ which are straightforward.

{ \small \input{code/concat} }
\vspace{1cm}
The image below visualises how the guide table speeds up
concatenation.
\begin{center}  
	\includegraphics[width=0.95\linewidth ]{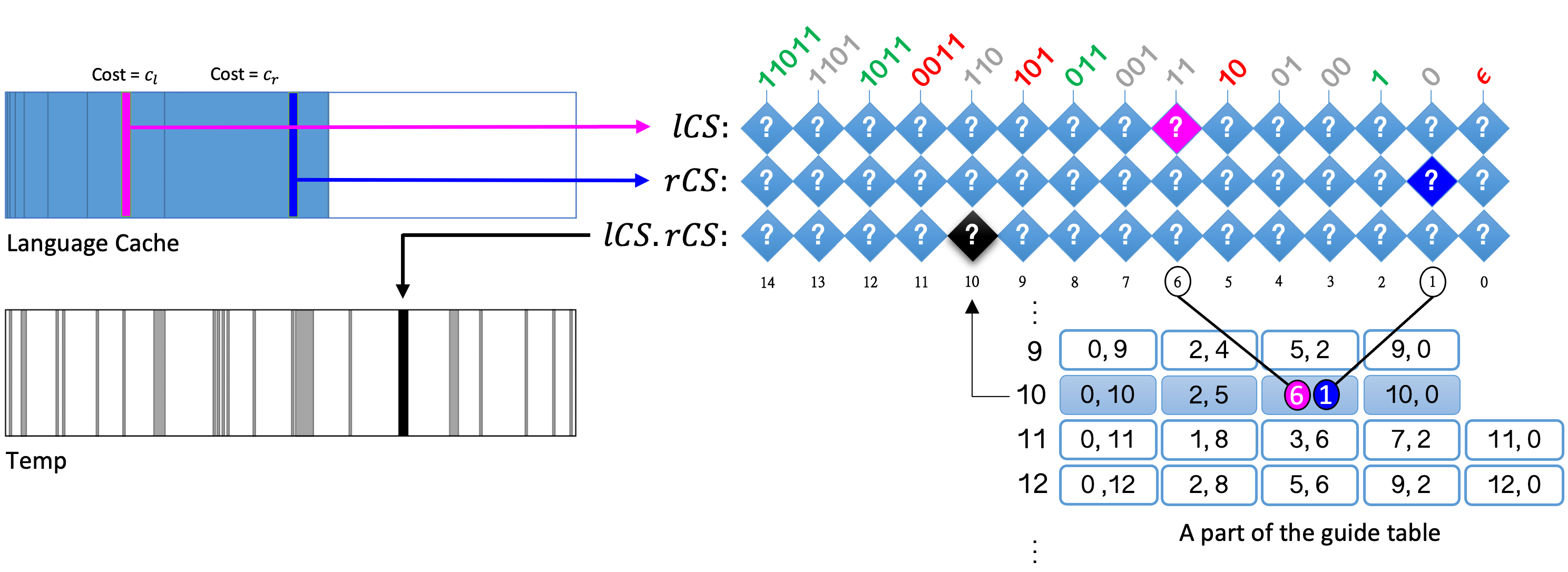}
\end{center}
 Checking if $w \in \IC{P \cup N}$ is accepted by $r_1 \cdot r_2$
 means checking, for all possible splits $w_1 \cdot w_2 = w$, if $w_i$
 is accepted by $r_i$.  This amounts to folding (in the sense of
 functional programming) over all splits.  As pairs $w_1, w_2$ are
 infixes of $w$, fast access to infixes is crucial, and the reason why
 our CS have domain $\IC{P \cup N}$ rather $P \cup N$.  If, given
 language cache entries $lCS$ and $rCS$, we want to compute $lCS \cdot
 rCS$, the CS representing concatenation of $lCS$ and $rCS$, we have
 to compute this fold for each word $w$---more precisely for each bit
 $k$ in the new bitvector $lCS \cdot rCS$.  The loop starting on Line
 \ref{concat:line:7} does this.  Each row in the guide table
 corresponds to, and is created as soon as $(P, N)$ is available, from
 words $w$. Each row in the guide table is a \EMPH{contiguous} list,
 hence bitvector, of pairs $(i, j)$ corresponding to a split $w = w_i
 \cdot w_j$. Here $i$ is the position of $w_i$ in CSs, and $j$ the
 position of $w_j$. (This indexing uses a power-of-two
 representation aka one-hot \cite{Wikipedia:onehot}, details
 omitted.)  The image highlights the word $w =$ ``110'' at index 10 in the
 language cache, and the guide table row for ``110''. The guide table
 entry $(6, 1)$ corresponds to splitting ``110'' into ``11'' and ``0''. Here
 $6$ is the index of ``11'' in each CS (read from $lCS$) while 1 is the
 index of   ``0'' (read from $rCS$).  Line \ref{concat:line:9}
 computes if this particular split generates $w$, and Line
 \ref{concat:line:10} computes the disjunction with other possible
 splits (we fold over all splits, as fast exits are data-dependent
 branching and problematic on GPUs).

\PARAGRAPH{Relationship between CPU and GPU implementation}
One purpose of the present work is to understand the speed-ups over
CPUs that can be gained by a GPU-friendly implementation of REI.
Alas, there was no suitable existing reference implementation, so we
had to produce one ourselves.  Ideally, we'd like to implement the
algorithm only once, and run it on both, CPU and GPU. But we found
CUDA programming in a GPU friendly manner leads to unnatural CPU code,
that would probably perform badly on a CPU.  So we implemented the
algorithm twice, once for CPUs and once for Nvidia GPUs.  Note that
many modern CPUs offer GPU-like features, \EG Intel's streaming
extensions \cite{strsime} or Arm's Scalable Vector Extension
\cite{NigelS:armscave}. We explicitly avoided using those, since they
are blurring the lines between CPUs and GPUs, making the comparison
less informative.

%% file: code/main.tex
\begin{algorithm}
	
	\scriptsize
	\caption{Main function of synthesis algorithm}\label{algo_pseudo_main2}
	
	\begin{algorithmic}[1]
		
		\State \textbf{Input} Positive and negative examples $\mathcal{(P, N)}$, $cost$, $maxCost$
		\State \textbf{Output} A minimal RE \WRT cost and $maxCost$  and consistent with $\mathcal{(P, N)}$, otherwise "not\_found"
		
		\\\hrulefill
		
 		\If {$\mathcal{P} == \{\}$} \Return $\emptyset$ \EndIf  \label{main:line:init:1}
 		\If {$\mathcal{P} == \{""\}$} \Return $\epsilon$ \EndIf				\label{main:line:init:2}
 		\State $languageCache$ = [list of CSs of alphabet] \Comment{$languageCache$ is  global variable} \label{main:line:3}
 		\For{$c \gets cost(\epsilon) + 1$ to $maxCost$}                 \label{main:line:4}
                   \State $questions$ = buildQuestionMark( $c$ - $cost(?)$)                 \label{main:line:5}
                   \State $stars$ = buildStar$( c - cost(*) )$ \label{main:line:6}
                   \State $concats$ = buildConcat$( c - cost(\cdot) )$ \label{main:line:7}
                   \State $unions$ = buildUnion$( c - cost(+) )$ \label{main:line:8}
                   \State $languageCache[c]$ = questions ++ stars ++ concats ++ unions \Comment{++ is concatenation} \label{main:line:9}		
 		\EndFor
                \State \Return "not\_found"	\Comment{Procedures in loop will return solution directly to caller of main, if found}
	\end{algorithmic}
\end{algorithm}

%% file: code/concat.tex
\begin{algorithm}
	
	\scriptsize
	\caption{Pseudocode for concatenation ($buildConcat$ procedure in \ALG \ref{algo_pseudo_main2})}\label{algo_pseudo_concat}
	
	\begin{algorithmic}[1]
		
		\State \textbf{Input} cost $c$, globals used: \texttt{languageCache}, P, N
		\State \textbf{Output} A list of new CSs generated by \textit{concatenation}
		
		\\\hrulefill
		
		\State outList $\gets$ [] \label{concat:line:1}
                \ForAll{L, R \text{such that} L + R = c}    \label{concat:line:2}
		\ForAll{$lCS \in languageCache(L)$}    \label{concat:line:3}
			\ForAll{$rCS \in languageCache(R)$}    \label{concat:line:4}
				\State $i \gets 1$   \label{concat:line:5}
				\State $newCS \gets 0$   \label{concat:line:6} 
				\For{$w \gets 0$ to $\CARD{\IC{P \cup N}} - 1$}   \label{concat:line:7}
					\ForAll{pair ($l$, $r$) $\in$ gt[$w$]}    \label{concat:line:8}
						\If{$(lCS\ \& \ l) \neq 0$ {\bf and} $(rCS\ \& \ r) \neq 0$}   \label{concat:line:9}
							\State $newCS \gets newCS\ | \ i$   \label{concat:line:10}
						\EndIf   \label{concat:line:11}
					\EndFor   \label{concat:line:12}
					\State $i \gets i \ll 1$   \label{concat:line:13}
				\EndFor   \label{concat:line:14}
				\State $isUnique \gets$ hashSet.insert($newCS$)   \label{concat:line:15}
				\If{$isUnique$}   \label{concat:line:16}
					\If{$newCS \models (P, N)$}   \label{concat:line:17}
						\State print $newCS$ and terminate program  \label{concat:line:18}
					\EndIf   \label{concat:line:19}
					\State outList.insert($newCS$)   \label{concat:line:20}
				\EndIf   \label{concat:line:21}
			\EndFor   
		        \EndFor   
                        \EndFor   
		\State \Return outList   \label{concat:line:25}
		
	\end{algorithmic}
\end{algorithm}

%% file: measurements.tex
\section{Evaluation of algorithm performance}\label{measurement}

\input{measurementsBackground}

\input{measurementsBenchGen}

\input{measurementsGPUonly}

\input{measurementsGPUCPU}

\input{measurementsCPUAR}
\input{measurementsConclusion}

%% file: measurementsBackground.tex
Contemporary ML  is an empirical field, and new
algorithmic approaches ought to be evaluated on reproducible
benchmarks.  To keep measurement and contributions focused, our
evaluation centres on the speed of our algorithm on a GPU. We'd like
to compare our work with existing comparable precise and minimal REI,
on widely agreed upon benchmarks.  This proved difficult: all existing
implementations of REI we consider compromise on
precision or minimality (often both).  Existing benchmarks
are unsuitable because they are either much too easy for \ALGO
 or they use large
alphabets and long strings that lead to out-of-memory errors in our
implementation.  (Other approaches that benchmark with large alphabets
and long strings compromise on precision, so solve a much easier
problem.)  In short: there is no comparable CPU implementation, and no
suitable benchmark suite. We solve both problems by implementing our
algorithm on a CPU and a GPU, and developing  suitable benchmarks.
All measurements and related artefacts necessary for
reproducing our measurement are available from
\cite{ArtifactGithub}. For brevity, the paper discusses only the most
interesting observations.

\subsection{Hardware and Software Used for Benchmarking}
Benchmarks in Sections \ref{measurement_cpu_vs_gpu} and
\ref{measurement_gpu_search_order} run on \textsc{Google Colab Pro}
\cite{GoogleColabPro}.  We use Colab Pro because it is a widely used
industry standard for running ML workloads.  Another reason is that we
did not have access to modern GPUs outside the cloud. \textbf{Colab CPU parameters:} 
Intel Xeon (``cpu family 6, model 79''), 2.20 GHz, RAM: 25
GB, running Ubuntu. We use the \texttt{g++} compiler, version 7.5.0, with
the \texttt{-O3} optimisation setting. We use \texttt{std::unordered\_set} \cite{CPPStandard:unoset}
to implement uniqueness checking. From now on we will refer to
this as Colab-CPU. \textbf{Colab GPU parameters:} Nvidia
A100-SXM4-40GB, Driver: Nvidia-SMI460.32.03, RAM: for comparison, we
restricted the program's memory usage to the 25 GB available on
the Colab-CPU, CUDA Version: 11.2.  We use the \texttt{nvcc} Nvidia
CUDA compiler driver, with CUDA compilation tools version 11.2.152.
We use the WarpCore library Version 1.0.0-alpha.1 \cite{JuengerD:warpcoreImplementation} to implement uniqueness checking.   From now on we will
refer to this as Colab-GPU. Benchmarks in \SEC
\ref{measurement_cpu_alpharegex} runs on a MacBook Pro.
\textbf{Laptop CPU parameters:}  with a 2.5 GHz Quad-Core Intel Core
i7, with 16 MB RAM.  We compile CPU C++ code using Apple clang version
11.0.3.  We compile \ALPHAREGEX using version 4.12.0 of the Ocaml
system. We compile with the native-code compiler (ocamlopt).  From now
on we will refer to this as Laptop-CPU.

\subsection{Threats to Validity}
Benchmarking is fraught with methodological difficulties that we are
intimately aware of, see
\cite{BarrettEdd:virmacwbhac,DehghaniM:benlot,HookerJN:tesheuwhiaw,GreggB:evaevaabc}
for a discussion.  A comparison between CPU and GPU is intrinsically
apples-to-oranges, and there are numerous ways in which our
measurements could be improved. 
\begin{itemize}

\item  It is unclear to what
  extent GPU, CPU on Google Colab Pro are virtualised. This may affect
  the reproducibility of measurements.

\item For the Colab-CPU we could neither determine the
  exact version of the processor nor the version of Ubuntu.

\item Our benchmarks are largely random strings, and those are likely
  quite different from strings that we expect to see in practically
  relevant REI.  We
  conjecture that random strings over an alphabet $\Sigma$ tend to be
  more difficult for grammar inference than more structured, human written examples.

\item  Benchmarking against \ALPHAREGEX compares a C++ with Ocaml, which could be seen as
  disadvantaging the \ALPHAREGEX. On the other hand, \ALPHAREGEX does not
  always return a minimal regular expression, so solves a 
  simpler problem.

\item We were  hampered by a ``measurement threshold'' of around 0.2
  seconds, a minimal time the Colab-GPU would take
  on \EMPH{any} task, including toy programs that do nothing at
  all on the GPU. We believe that this might be GPU
  latency \cite{WilperH:undvisooalinns}, possibly compounded by the
  Colab framework. This stood in the way of evaluating the performance
  of \ALGO on small benchmarks.

\end{itemize}
 We believe the speedups we find are unlikely only the effect of
 measurement bias. We encourage
 others to replicate our experiments and
 improve our measurement methodology.

%% file: measurementsBenchGen.tex
\subsection {Benchmark Construction}
A good benchmark suite should be  tunable by a small number of
explainable parameters that allows users to achieve hardness levels, from
trivial to beyond the edge-of-infeasibility, and any point in-between.
The benchmarks should be suitably random to reduce biasing 
measurements, yet remain fully reproducible.  We are interested in
both, space and time complexity, since
our algorithm make a space-time trade-off, and quickly solves
virtually every problem instance that fits in the available memory. That
means we need to be able to fine-tune benchmarks to target memory
availability.  \ALGO's memory usage is governed by the size of $\IC{P \cup
  N}$, which, in turn depends on two related factors: the length of
the longest strings in $P \cup N$, and what might be called the
heterogeneity of infixes: \EG $\IC{\{aaa, aa\}} = \{aaa,
aa, a, \epsilon\}$ is  smaller than $\IC{\{abc, de\}}
= \{abc, ab, bc, de, a, b, c, d, e, \epsilon\}$, despite both being
computed from two strings of identical lengths.  This suggest a
reproducible way of constructing benchmarks with the following natural
parameters.
  \begin{itemize}

  \item  Alphabet $\Sigma$,
    \item $le$ is the
      maximal  length of example strings,
      \item $p$ and $n$, the numbers of positive and negative
        examples, respectively.

  \end{itemize}
With those parameters, we define two complementary benchmark
generation schemes. Both create instances $(P, N)$ by sampling
uniformly from two different spaces of random strings.
  \begin{itemize}
    
  \item \textsc{Type 1:} 
  $
   \{ (P, N) \in \Sigma^{\leq le} \times \Sigma^{\leq le}\ |\ \forall w \in P \cup N. \CARD{P} = p, \CARD{N} = n, P \cap N = \emptyset \}
  $

 \item \textsc{Type 2:} 
  $
  \{ ((P_0, ..., P_{le}), (N_0, ..., N_{le})) \in Y \times Y\ |\ \Sigma_i \CARD{P_i} = p, \Sigma_i \CARD{N_i} = n,  \forall i. P_i \cap N_i = \emptyset \}
  $

\end{itemize}
Here $Y$ is $\POWERSET{\Sigma^0} \times \cdots \times
\POWERSET{\Sigma^{le}}$, and the benchmark corresponding to $((P_0,
..., P_{le}), (N_0, ..., N_{le}))$ is $(\bigcup_{i}P_i,
\bigcup_{i}{N_i})$. \textsc{Type 1} and \textsc{Type 2} have different flavour: since
there are exponentially more long strings than short, \textsc{Type 1}
specifications are dominated by long strings. In order also to be able
to study the effects of short strings in specifications, \textsc{Type 2},
gives each length the same chance of occurring in positive
or negative examples.  Hence short strings, like $\epsilon$, are
likely to be in most \textsc{Type 2} specifications.  Our main 
benchmarks, used below, are generated using the following
parameter.  The alphabet is $\{0, 1\}$. The remaining
parameters have been chosen to be as hard as possible on the GPU while
avoiding running out of memory.  \textsc{Type 1} benchmarks: $p$ and $n$ both
range over 8 to 12, and $le$ over 0 to 7.  \textsc{Type 2} benchmarks: $p$ and
$n$ both range over 7 to 14, and $le$ over 0 to 10.  While \ALGO can
deal with arbitrary alphabets (with the expected increase in search
space size), we restrict our attention to $\{0, 1\}$ because our main
point of comparison, \ALPHAREGEX, can only handle binary alphabets.

%% file: measurementsGPUonly.tex
\begin{FIGURE}
	\includegraphics[width=1\linewidth]{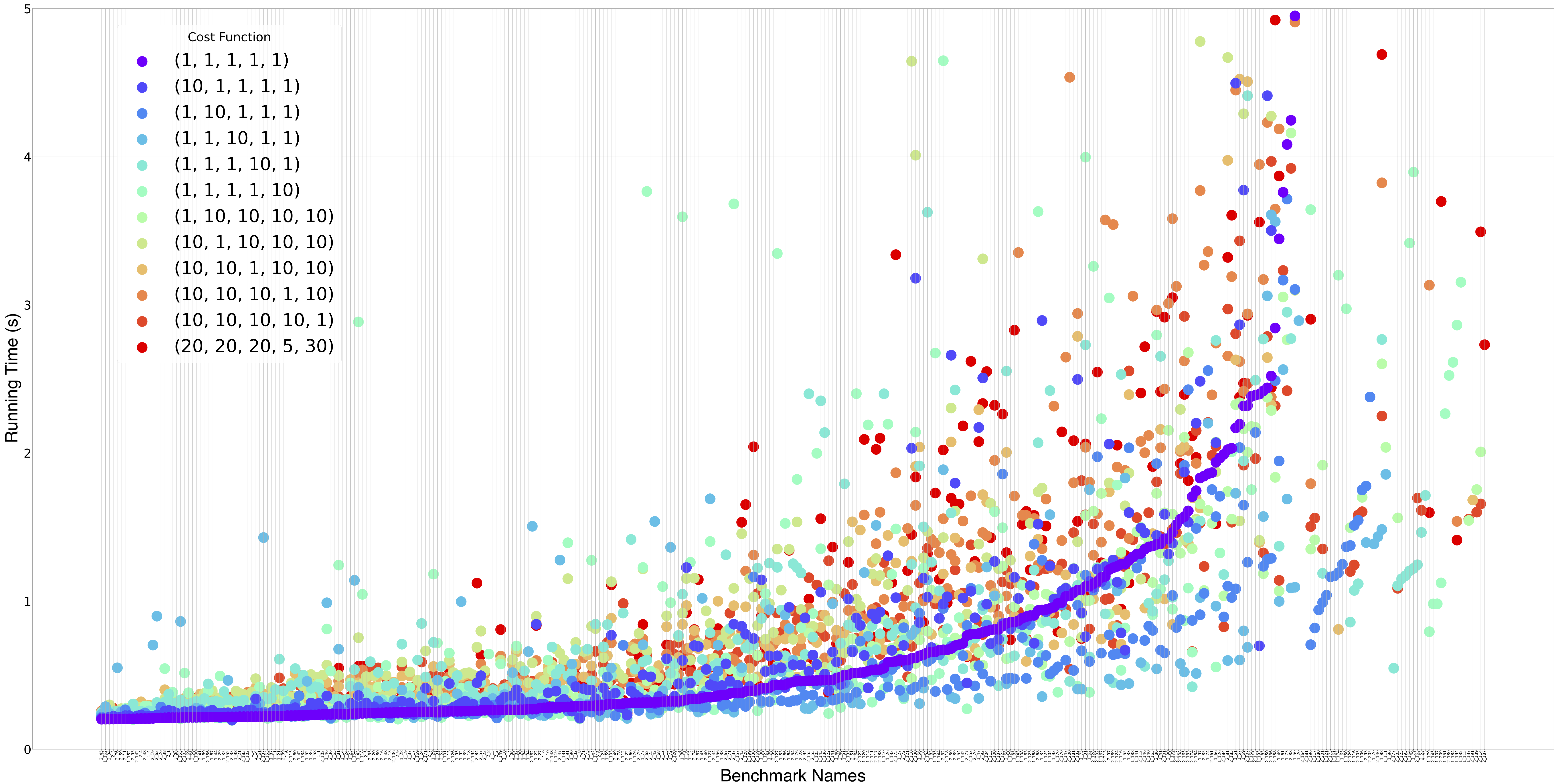}
	\caption{Plotting 3325 benchmarks with 12 different cost functions on the Colab-GPU.
		The x-axis are the benchmark names, sorted by
		increasing duration of each named benchmark using the $(1, 1, 1,
		1, 1)$ cost function. Benchmarks that run more than 5 seconds (only 3.62\%), or run out of memory,
		are omitted.}\label{figure_type1-2_comparison}
\end{FIGURE}

\PARAGRAPH{Measurement (1): impact of cost functions}\label{measurement_gpu_search_order}
How big is the impact of cost function on run time? Intuitively, it
should be strong, since any specific cost function induces a search
order, and the solution to a given synthesis problem can come earlier
or later in the search process depending on that order. Figure
\ref{figure_type1-2_comparison} hones in on, and quantifies the effect
of search order on synthesis performance. The measurements were done
with the Colab-GPU. We run and time 5160 benchmarks (200 examples from 
\textsc{Type 1}, and 230 from \textsc{Type 2}, all with 12 different cost functions), on the 
Colab-GPU 3 times, and take the average  of those 
three runs for each benchmark. We plot those examples that don't timeout 
within 5 seconds at least for one of the cost functions).  We do
\EMPH{not} run the measurements on a CPU because that would be too
time-consuming (order of weeks). Since the algorithm is the same on
CPU and GPU, and Table \ref{table_gpu_cpu_comparison} indicates
that, indeed, search order affects CPU and GPU similarly, we
conjecture that we would see similar effects with CPUs.  Here is a
summary of observations.
\begin{itemize}

\item Measurements cluster on the bottom left of the figure, meaning
  that slow benchmarks are the exception rather than the rule: 60\% of
  benchmarks run in under 1 second and 73\% under 2
  seconds. We discuss outliers briefly at the end of this section.

\item The (1, 1, 1, 1, 1) cost function shows a clean and steep
  increase in synthesis time, which we do not see for  
  other cost functions. This is partly a consequence of the fact that
  the x-axis is benchmarks, ordered by increasing time using (1 1 1
  1 1) (details of sorting in case of tie are in \cite{ArtifactGithub}).
  
\item The (1, 1, 10, 1, 1) cost function which makes the
  Kleene-star expensive, is often fast. This might be surprising since
  the Kleene-star is the one mechanism regular expressions have to
  'generalise', to exploit patterns.  Our benchmarks use random
  strings, so it's unlikely that we get many opportunities that make
  the Kleene-star useful. Therefore cost functions that avoid the cost
  of the Kleene-star are likely to be faster. It would be
  interesting to use this cost function on benchmarks with a lot of
  repetition.
  
\item The (1, 1, 1, 1, 10) cost function that makes union expensive is
  usually the slowest one. We can infer from long running-times that
  the algorithm does not run out of memory.  This is only possible if most
  newly generated CSs fail their uniqueness checks.

\end{itemize}

%% file: measurementsGPUCPU.tex
\begin{TABLE}
	\caption{Comparison of \ALGO on hardest examples, using the
	  Colab-CPU and Colab-GPU. }\label{table_gpu_cpu_comparison}
	\input{data/table_gpu_cpu_comparison}        
\end{TABLE}

\PARAGRAPH{Measurement (2): CPU vs GPU versions of our algorithm}\label{measurement_cpu_vs_gpu}
Table \ref{table_gpu_cpu_comparison} compares \ALGO on a CPU and a
GPU. Our choice of benchmark parameters resulted in benchmarks that
can fit into Colab-GPU, but take on average about 1 hour each on the
Colab-CPU.  For each pair of (type, cost-function), we chose from the
previous benchmark the longest-running benchmark that neither ran
out-of-memory nor timed out. We run and time these $24$ benchmarks 3
times, and report the average  of those three
runs for each benchmark.  Here is a summary
of core observations.
\begin{itemize}

\item Our GPU is three orders of magnitude faster than our CPU version
  of the same algorithm, and the speed-up does not depend on the chosen
  cost function.
 
\item Almost half of \textsc{Type 1} examples in Table
  \ref{table_gpu_cpu_comparison}, contain $\epsilon$. This is
  surprising, because  \textsc{Type 1} strongly favours long strings. Indeed, we created \textsc{Type 2} only
  so that we can also run benchmarks with specifications contain short
  strings, including, in particular $\epsilon$. So $\epsilon$ seems to
  make inference disproportional harder. We conjecture that this is
  an effect of using regular expressions, rather
  than an artefact of our algorithm.
  
\end{itemize}

%% file: data/table_gpu_cpu_comparison.tex
\small
\begin{tabular}{|ccccc|c|ccc|}
	\toprule
	
	\multicolumn{5}{c}{\textbf{Input}} & \multicolumn{1}{c}{\textbf{CPU}} & \multicolumn{3}{c}{\textbf{GPU}} \\
	
	\cmidrule(rl){1-5} \cmidrule(rl){6-6} \cmidrule(rl){7-9}
	
	Type & No & \CARD{P} & \CARD{N} & Cost Function & Sec & Sec & Speed-up & \CARD{REs} \\

	\midrule

	1 & 50 & 10 & 12 & (1, 1, 1, 1, 1) & 5080.7850 & 4.9512 & \textbf{1026x} & 26,774,099,142 \\
	1 & 51 & 12 & 9 & (10, 1, 1, 1, 1) & 4699.8137 & 4.4966 & \textbf{1045x} & 23,824,118,297 \\
	1 & 73 & 10 & 11 & (1, 10, 1, 1, 1) & 5805.2168 & 3.7144 & \textbf{1562x} & 22,703,639,676 \\
	1 & 20 & 9 & 9 & (1, 1, 10, 1, 1) & 2893.4835 & 2.8935 & \textbf{1000x} & 13,567,472,188 \\
	1 & 73 & 10 & 11 & (1, 1, 1, 10, 1) & 2901.9297 & 2.9504 & \textbf{983x} & 11,706,686,339 \\
	1 & 31 & 8 & 9 & (1, 1, 1, 1, 10) & 5856.6925 & 3.9973 & \textbf{1465x} & 14,210,157,835 \\
	1 & 57 & 12 & 10 & (10, 10, 10, 10, 1) & 2804.6793 & 3.4322 & \textbf{817x} & 14,163,906,090 \\
	1 & 50 & 10 & 12 & (10, 10, 10, 1, 10) & 4519.9456 & 4.9096 & \textbf{920x} & 23,349,552,935 \\
	1 & 57 & 12 & 10 & (10, 10, 1, 10, 10) & 4301.8548 & 4.5243 & \textbf{950x} & 20,257,045,497 \\
	1 & 97 & 12 & 12 & (10, 1, 10, 10, 10) & 5608.7286 & 4.7782 & \textbf{1173x} & 19,680,542,658 \\
	1 & 61 & 12 & 10 & (1, 10, 10, 10, 10) & 2915.0938 & 3.0532 & \textbf{954x} & 14,322,039,866 \\
	1 & 88 & 12 & 9 & (20, 20, 20, 5, 30) & 6899.0045 & 4.6904 & \textbf{1470x} & 25,193,577,825 \\
	2 & 88 & 14 & 8 & (1, 1, 1, 1, 1) & 3783.9772 & 4.2462 & \textbf{891x} & 23,697,549,545 \\
	2 & 150 & 14 & 12 & (10, 1, 1, 1, 1) & 4228.2773 & 4.4120 & \textbf{958x} & 23,125,803,623 \\
	2 & 158 & 12 & 14 & (1, 10, 1, 1, 1) & 2975.9956 & 2.4887 & \textbf{1195x} & 11,432,891,412 \\
	2 & 136 & 11 & 14 & (1, 1, 10, 1, 1) & 3374.8873 & 3.6080 & \textbf{935x} & 18,241,755,827 \\
	2 & 107 & 12 & 12 & (1, 1, 1, 10, 1) & 2432.4320 & 4.4120 & \textbf{551x} & 24,954,272,802 \\
	2 & 32 & 10 & 7 & (1, 1, 1, 1, 10) & 7400.8135 & 4.6482 & \textbf{1592x} & 16,729,795,052 \\
	2 & 136 & 11 & 14 & (10, 10, 10, 10, 1) & 2907.9182 & 3.9689 & \textbf{732x} & 17,476,988,322 \\
	2 & 200 & 13 & 8 & (10, 10, 10, 1, 10) & 9687.7952 & 4.5366 & \textbf{2135x} & 6,037,014,423 \\
	2 & 107 & 12 & 12 & (10, 10, 1, 10, 10) & 3383.1937 & 4.5071 & \textbf{750x} & 20,697,274,025 \\
	2 & 81 & 8 & 14 & (10, 1, 10, 10, 10) & 3497.9013 & 4.6699 & \textbf{749x} & 21,869,903,022 \\
	2 & 88 & 14 & 8 & (1, 10, 10, 10, 10) & 3405.5536 & 4.1602 & \textbf{818x} & 21,889,508,744 \\
	2 & 158 & 12 & 14 & (20, 20, 20, 5, 30) & 5804.8112 & 4.9228 & \textbf{1179x} & 23,163,079,580 \\
	
	\midrule
	
	\multicolumn{5}{c}{\textbf{Average}} & \multicolumn{1}{c}{4465.4493} & \multicolumn{1}{c}{4.1238} & \multicolumn{1}{c}{\textbf{1077x}} & \multicolumn{1}{c}{19,127,861,447} \\
	
	\bottomrule
\end{tabular}

%% file: measurementsCPUAR.tex
\begin{TABLE}
	\caption{Running \ALGO and \ALPHAREGEX on the laptop-CPU.
          Benchmarks labelled with  $\dagger$ replace the ``wild
          card'' $x$ with the intended $0 + 1$, since \ALGO does not
          support their ``wild card'' heuristic.  Cost 
          \textbf{\underline{bold and underlined}} is not minimal.  On
          the Colab-GPU, \ALGO does not accept no6 and no9, because
          they require 128 and 256 bits for their respective $\IC{P
            \cup N}$, which WarpCore does not currently support; no9
          needs > 256 bits which even our CPU version does not
          currently support. All but three examples that we can run on
          the Colab-GPU, finish in approx.~0.2 seconds, \IE below the
          measurement threshold.}\label{table_gpu_lee_comparison}
\input{data/table_lee_comparison}        
\end{TABLE}

\PARAGRAPH{Measurement (3): comparison with \ALPHAREGEX}\label{measurement_cpu_alpharegex}
We compare our algorithm against \ALPHAREGEX, the
state-of-the-art REI system.
We use  the Laptop-CPU because we could not get
Ocaml, required for \ALPHAREGEX, to run on the Colab-CPU.  We do not
run our own benchmarks because, after informal experiments, we felt \ALPHAREGEX 
would take too long. Moreover, most of our benchmarks contain $\epsilon$, which \ALPHAREGEX
does not handle.
Instead we use the benchmark from
\cite{LeeM:synregefefiaaCODE,LeeM:synregefefiaa} (slightly adapted):
because \ALPHAREGEX solves those quickly.
We run these benchmarks only use the CPU version of
\ALGO since almost all are solved on the Colab-GPU by \ALGO below the
measurement threshold.  We adapt some benchmarks because \ALPHAREGEX
uses the ``wild card'' heuristic, which \ALGO does not support: we
replace 'X' by $0+1$, the meaning intended by \ALPHAREGEX.  In order to
mitigate the methodological problems arising from comparing C++ and
Ocaml, we report the regular expressions
checked by both for compliance with the ambient specification, as those depend only on
the algorithm. Here is a
summary of observations.
\begin{itemize}

\item \ALPHAREGEX does not always return minimal-cost regular
  expressions in nearly 25\% of their own benchmarks. This is surprising, given
  the abstract of \cite{LeeM:synregefefiaa}. We believed this is a
  direct consequence of their heuristics\footnote{In passing we note
  that sometimes the lower-cost regular expression we synthesise does
  not meet the corresponding English language description in
  \cite{LeeM:synregefefiaa}. For example, in benchmark ``no25'', the
  description in English is ``at most one pair of consecutive
  1s''. We synthesise $0+((1+00)(0+1))^{*}$, which is lower cost than the
  solution \ALPHAREGEX finds, and meets their all positive and negative
  examples, but accepts strings like 1111.}.

\item \ALPHAREGEX's pruning heuristics often work well, and can
  sometimes decrease the number of regular expressions checked by an
  order or two of magnitude. Surprisingly, and despite their pruning
  heuristics, in about 20\% of benchmarks \ALPHAREGEX checks more
  regular expressions than \ALGO. In all benchmarks, \ALGO is faster
  despite generating and checking many more regular
  expressions. 

\item \ALPHAREGEX is always slower than the CPU version of \ALGO, in
  an extreme case by more than three orders of magnitude. 
  
\item The benchmark running out-of-memory with \ALGO (``no9'') can be
  executed by \ALPHAREGEX, albeit not within 20000
  sec\footnote{\ALPHAREGEX solves it quickly using the ``wild card''
  heuristic.}.
  
\end{itemize}

%% file: data/table_lee_comparison.tex
\newcommand{\XXX}[1]{\ALPHAREGEXREPO/blob/master/benchmarks/#1}
\newcommand{\YYY}[1]{\AREPO/results/alpharegexComparison//#1}
\small
\begin{tabular}{|l|ccc|cc|ccc|}
	
	\toprule
	
	\multicolumn{1}{c}{\textbf{}} & \multicolumn{3}{c}{\textbf{Running Times (sec)}} & \multicolumn{2}{c}{\textbf{Cost(RE)}} & \multicolumn{3}{c}{\textbf{\CARD{REs}} }  \\
	
	\cmidrule(rl){2-4} \cmidrule(rl){5-6} \cmidrule(rl){7-9}
	
	No & \AR  & \ALGOSHORT & Speed-up & \AR  & \ALGOSHORT & \AR  & \ALGOSHORT & Increase\\
	
	\midrule
	
	no1$^{\dagger}$ & 0.0092 & 0.0054 & \textbf{2x} & 85 & 85 & 27 & 77 & 2.85x\\
	no2$^{\dagger}$ & 0.0402 & 0.0042 & \textbf{10x} & \textbf{\underline{155}} & 110 & 704 & 470  & 0.67x\\
	no3$^{\dagger}$ & 52.6748 & 24.1921 & \textbf{2x} & 280 & 280 & 282,815 & 32,329,412 & 114.31x\\
	no4$^{\dagger}$ & 0.0277 & 0.0051 & \textbf{5x} & \textbf{\underline{155}} & 140 & 520 & 3,821    & 7.35x \\
	no5$^{\dagger}$ & 50.5782 & 3.144 & \textbf{16x} & 265 & 265 & 292,115 & 9,960,260  & 34.10x \\
	no6$^{\dagger}$ & 231.4295 & 4.1491 & \textbf{56x} & 240 & 240 & 659,702 & 4,731,056   & 7.17x \\
	no7 & 7.211 & 0.5716 & \textbf{13x} & \textbf{\underline{230}} & 225 & 92,702 & 581,659  & 6.27x \\
	no8 & 0.0673 & 0.0096 & \textbf{7x} & 160 & 160 & 1,318 & 7,776   & 5.90x \\
	no9$^{\dagger}$ & >20000 & N/A  & \textbf{N/A} & N/A & N/A & N/A & N/A& N/A \\
	no10 & 1.4267 & 0.0057 & \textbf{250x} & 155 & 155 & 21,457 & 6,772  & 0.32x \\
	no11 & 0.01 & 0.0041 & \textbf{2x} & 85 & 85 & 92 & 77  & 0.84x\\
	no12 & 0.1614 & 0.0137 & \textbf{12x} & 185 & 185 & 3,721 & 23,675  & 6.36x \\
	no13 & 1.105 & 0.1961 & \textbf{6x} & 220 & 220 & 17,957 & 483,161 & 26.91x \\
	no14$^{\dagger}$ & >20000 & 116.5046 & \textbf{N/A} & N/A & 310 & N/A & 261,293,189 & N/A\\
	no15$^{\dagger}$ & 25.9066 & 0.2952 & \textbf{88x} & 240 & 240 & 187,484 & 1,721,174 & 9.18x\\
	no16$^{\dagger}$ & 3.8588 & 0.0594 & \textbf{65x} & 205 & 205 & 32,039 & 225,377  & 7.03x\\
	no17 & 2.1345 & 0.326 & \textbf{7x} & \textbf{\underline{230}} & 225 & 31,476 & 659,386 & 20.95x\\
	no18 & 0.1004 & 0.0074 & \textbf{14x} & 155 & 155 & 1,710 & 8,010  & 4.68x\\
	no19$^{\dagger}$ & 0.0207 & 0.0057 & \textbf{4x} & 130 & 130 & 164 & 1,867  & 11.38x\\
	no20$^{\dagger}$ & 25.6899 & 0.0192 & \textbf{1338x} & 160 & 160 & 97,510 & 21,135  & 0.22x\\
	no21 & 0.8064 & 0.1993 & \textbf{4x} & 225 & 225 & 18,887 & 481,762  & 25.51x\\
	no22$^{\dagger}$ & 68.6433 & 1.1519 & \textbf{60x} & 265 & 265 & 495,783 & 6,202,349  & 12.51x\\
	no23 & 1.5421 & 0.028 & \textbf{55x} & \textbf{\underline{210}} & 180 & 22,411 & 32,068 & 1.43x\\
	no24 & 10.7292 & 0.0491 & \textbf{219x} & 200 & 200 & 127,893 & 109,433  & 0.86x\\
	no25 & 19.182 & 3.7642 & \textbf{5x} & \textbf{\underline{265}} & 240 & 260,104 & 3,205,741  & 12.32x\\
	
	\bottomrule
	
\end{tabular}

%% file: measurementsConclusion.tex
\PARAGRAPH{A note on outliers}
We have repeatedly stated that \ALGO solves virtually all benchmarks
(that fit into memory at all) in a few seconds. However, there is a
small number of outliers that take much longer. The table below
quantifies outliers \WRT the full benchmark suite.

{\small
\begin{center}
\begin{tabular}{ |c|c|c|c|c|c|c|c|c|c|c|c| } 
  \hline
  Duration (sec) & <2 & <3 & <4 & <5 & <10 & <25 & <50 &  <100 & <200 & <400 & <800 \\ \hline
 \% of benchmarks  & 89.48 & 94.06 & 95.71 & 96.38 & 98.14 & 98.84 & 99.28 & 99.59 & 99.83 & 99.91 & 100.00 \\
 \hline
\end{tabular}
\end{center}
}

\PARAGRAPH{Performance evaluation}
\ALGO  is exponential in (asymptotic)  space and time complexity.
\ALGO terminates no later than with the maximally overfitted regular expression
 $w_1 + ... + w_i$ for $(P, N)$, assuming $P = \{w_1, ..., w_i\}$, see
(\ref{running_example_132_overfit}) from the introduction. Let $X$ be
the number of regular expressions with  cost not exceeding that of $w_1 +
... + w_i$, then $X$ is an upper bound on the number of generated CSs.
Regarding space complexity, we store only unique
CSs. Let $Y$ be the number of unique CSs in the language cache, so
$Y \leq X$. Each CS uses approx.~$k$ bits where $k$ is size of
$\IC{P \cup N}$. We use additional memory for uniqueness checking and
reconstructing concrete regular expressions from CSs: overall
approx.~$3 \cdot k$ bits for each CS. This bound works regardless of
alphabet size. Hence the worst-case size of the language cache is
$3 \cdot k \cdot Y \leq 3 \cdot k \cdot X$ bits.    Getting tighter average and worst-case bounds on the
number  of unique CSs for a given specification is an interesting
open problem.  We leave a detailed investigation of the performance
overheads of cache misses, data-depended branching and hardware
synchronisation overhead (\EG insertion of unique CSs into the shared
language cache) as future work.

\PARAGRAPH{Summary of evaluation} We believe,
that, despite the difficulties with measurement methodology we noted,
the speed-ups we are seeing from running \ALGO, especially on GPUs, are
not just an effect of measurement bias. We believe that the main
reason for the performance improvement we are seeing, is that
our algorithm is  GPU-friendly.

%% file: conclusion.tex
\section{Conclusion}\label{conclusion}

\input{related}
\input{future}

%% file: related.tex
The present work combines three main themes: ML, GPU
programming and algorithms \& data structures. Each is vast and we
could not possibly do justice to those fields here. Instead, we
highlight some key works that have influenced our thinking.

\subsection{Related Work}

\PARAGRAPH{Program synthesis techniques}
FlashFill \cite{GulwaniS:autstrpisuioe} reinvigorated program
synthesis a decade ago.  While FlashFill does not do REI but
synthesises string transformers, there is conceptual overlap with
\ALGO, in that both return minimal solutions \WRT a cost function
($\mathtt{size}(\cdot)$ in FlashFill), and that both make crucial use of
infixes.  The main
differences are: (i) FlashFill represents infixes of strings with
a syntactic construct \EG $\mathtt{SubStr(s,2,5)}$, resp.~a DAG, and
explicitly represents start- and end-positions of infixes as numbers.
In contrast, infixes are
implicitly represented as bits in our CS and accessed by position.
(ii) \ALGO runs on GPUs, while, to the best of our knowledge,
FlashFill is implemented only for CPUs. We don't believe the FlashFill
algorithm can be implemented efficiently on GPUs without substantial
re-engineering.  (iii) Unlike our use of semirings which immediately
also generalise to context-free, and, indeed, all formal languages,
FlashFill's data structures are not related to abstract mathematics,
making it difficult to see how they generalise.  (iv) FlashFill's  cost function is not configurable.
(v) FlashFill is used as an incremental synthesis tool in Microsoft Excel. \ALGO is currently
not incremental. 

Our CSs are  a variation on the theme of \EMPH{observational
  equivalences}, a standard technique to mitigate the cost imposed on
synthesis by the redundancies of syntax: intuitively, programs are 
equivalent if they have the same behaviour in all contexts, \EG $r^*$ and $\epsilon +
r^*r$. An ideal synthesis mechanism  searches over canonical
representatives of programs quotiented by this equivalence, alas observational
  equivalence is not computable in general.  When
synthesising programs from examples, it is natural to consider
programs equivalent if they relate to the examples in the same manner,
FlashFill \cite{GulwaniS:autstrpisuioe} and TRANSIT
\cite{UdupaA:traspepwcs} do this. \ALGO does something subtly different: we
do not identify regular expressions \WRT to words in the examples, but instead
over-approximate and identify them if they have the same CS over the
infix-closure of the examples.
This over-approximation is vital for fast
bottom-up synthesis of regular expressions.  Other work instead
under-approximates: this is a heuristic for quickly
discarding obviously unsuitable candidates, \EG the ``fingerprints''
in superoptimisation \cite{BansalS:autgenops}.  Similar techniques are
used in \EG equality saturation \cite{NandiC:rewruliues}.

\PARAGRAPH{Regular expression synthesis from examples}
\ALPHAREGEX\ \cite{LeeM:synregefefiaa,LeeM:synregefefiaa2} works from positive and negative
examples, and a configurable (albeit only by editing and recompiling source code)
cost homomorphism. \ALPHAREGEX uses top-down, exhaustive search over
regular expressions extended with a concept of 'hole'. The clean
mathematical semantics of extended regular expressions enables elegant
pruning heuristics.  All synthesised regular expressions are precise.
\cite{LeeM:synregefefiaa} claims that 
synthesised regular expressions are minimal (``\EMPH{the method automatically
synthesizes the simplest possible regular expression that ...}'' ),
but we found several counterexamples (see
Table \ref{table_gpu_lee_comparison}). \ALPHAREGEX has two main
restrictions: binary alphabet only, and examples must not contain the
empty string (\ALGO has neither restriction).
\textsc{FlashRegex} \cite{LiY:flaregdarrfe} presents an interesting twist on
regular expression synthesis: regular expressions are often used in
security sensitive applications, and, if chosen naively, can enable
denial-of-service attacks, called ReDOS \cite{Wikipedia:redos}.
\textsc{FlashRegex} optimises for
generating regular expressions from positive and negative examples
that are not susceptible to ReDOS. Since this work is optimising in a
different direction from ours, the two approaches are not directly
comparable. However, an interesting research problem is to investigate
if \ALGO can be extended so it guarantees lack of ReDOS vulnerability
in addition to, or instead of minimality.

  A widespread use of regular expressions is for information
  extraction: $ \EXTRACT{r}{w} $ returns all substrings $w'$ of $w$
  such that $w' \in \LANG{r}$.  Examples include extracting URLs from
  a web-page. Much research has been done  on inference of regular expressions for  information
  extraction, often in the context of XML, or linguistics. We mention
  only \cite{BartoliA:autsynorefe,BartoliA:infregeftefe,LiY:regexplfie,BexGJ:infconread},
  where the reader can find more references.  The
  learning problem in those papers is the same as ours: from given
  positive and negative examples, construct a suitable regular
  expression. The resulting regular expression is not guaranteed to be
  minimal and it also not, in general, precise. The papers use genetic
  programming as search mechanism.

Often input to REI changes only gradually and incrementalising this
problem is an appealing proposition. Several papers 
consider this problem. We mention two:
  \cite{PanR:autrepore} design a heuristic algorithm called \textsc{RFixer}
  that repairs a regular expression with given positive and negative
  examples. When the regular expression is incorrect on the
  examples, RFixer automatically synthesises the syntactically
  smallest repair of the original regular expression that is correct
  on the given examples.  This can be seen as a  incremental
  regular expression synthesis.
  \cite{WangX:fidfilsdue} considers the widespread use of
  regular expressions for data filtering, for example to 'zoom in' on
  relevant data in a spreadsheet. This can be done with regular
  expressions: ``show me only the strings that start with something
  matching regular expression $r$''. Since construction of suitable
  regular expressions is considered hard, it should be learned from
  examples. Like us \cite{WangX:fidfilsdue} consider positive and
  negative examples, but they consider only the star-free subset of
  regular expressions, making the search space much easier.  We can
  already search in the star-free fragment, by setting $\COST{*}$ 
  high enough, however, our algorithm is not incremental.  In either
  case, incrementality typically requires sophisticated support data
  structures to work well. We leave the question of incrementalising
  our algorithm as important future work.
  

We are reluctant to compare with deep-learning based regular
expression synthesis, because existing implementations
like \cite{LiY:traregmmresbgar,LocascioN:neugenorefnlwmdk,ZhongZ:semregasbafgrefnls,ParkJU:sofreggrfnldusre}
all work from natural language specification and make neither
minimality nor precision guarantees. Indeed, given the intrinsic
ambiguity of natural language, what would such guarantees even mean?
In order to fill this notable lacuna, the present authors are
preparing a comparison between \ALGO and REI on generative
pre-trained transformers \cite{VaswaniA:attallyn}.

\PARAGRAPH{Acceleration of regular expression contains-checking}
Regular expressions are widely used, and performance-critical for many
applications.  So it is not surprising that there is work on
accelerating regular expressions with GPUs or even dedicated hardware.
It is crucial to understand that this existing work is accelerating a
subtly different problem, which we call the \EMPH{contains-check} for
regular expressions (also known under different terms, including, but
not limited to matching, pattern matching, evaluation, language
containment). The table below summarises the difference.

\begin{center}
\begin{tabular}{|l||l|l|}
\hline
          & REI & Contains-check \\ \hline\hline
   Input  & Sets of strings  & String         \\ \hline
   Output & Regular expression           &  True/False    \\ \hline
\end{tabular}
\end{center}

It is not clear that accelerating  REI is automatically also
advantageous for contains-checking, or vice-versa.  Be that as it may,
we briefly survey existing work on the acceleration of regular
expression contains-checking.

Much has been written about 
 implementing fast regular expression contains-checking on CPUs. We
mention only \cite{QiuJ:scafsmpvpfahos}, which presents two ideas that might be interesting
 on GPUs, too:  
the problem with contains-checks, from the point of view of
parallelism, is that the contains-check appears to be 
sequential. In order to parallelise contains-checking of  string $w$,
we could break $w$ into parts, \EG to check if $r_1 r_2$ contains
 $w$, we could break $w$ into  $w = w_1w_2$
and then check in parallel if $r_1$ contains $w_1$ and $r_2$ contains
$w_2$. The problem is that's it's not easy to know where to split $w$.
\cite{QiuJ:scafsmpvpfahos} suggests doing this speculatively, and perform a 'rollback'
in case of miss-speculation.
Contains-check acceleration on GPUs is a relatively new field, not
surprising given that GPUs are relatively recent and hard to program.
The first work to do so was \cite{CascaranoN:infnfapmogd}.  The paper
exploits the parallelism GPUs offer through a non-deterministic
automata representation of regular languages.
Later, \cite{ZuY:gpubnifmehsrem,LiuH:whygpuasaenahtmtf}
improves \cite{CascaranoN:infnfapmogd} with more carefully designed
data structures.

%% file: future.tex
\subsection{Future Work}\label{future}

It is natural to consider generalisation of the \ALGO algorithm to
other classes of grammars, whether more expressive (\EG context-free
or context-sensitive), or less (\EG restricted star-height).  Many of
our core algorithmic choices, in particular the choice of bitvectors to
represent languages, and the use of uniqueness as generic pruning
technique, work for any grammar.

It is also natural to ask if \ALGO can be made less memory intensive
by compromising on minimality or precision.  In other words, can
local search benefit from ideas for GPU acceleration of
global search?  In closing, we
sketch how to implement \EMPH{REI with error}, a simple local search
technique that requires changing only a few lines of code.
Intuitively it is clear that REI becomes easier if we drop the
requirement that the result be precise. Let's introduce an
\EMPH{allowed error}
parameter, a number between 0 and 1, that quantifies the
\EMPH{allowed error}. Solving $(P, N)$ then requires finding $r$ with
$r \models (P', N')$ such that $P' \subseteq P$ and $N' \subseteq N$,
and the allowed error is an upper bound on the fraction of $P\cup N$ that is not in $P' \cup N'$.
Consider the specification $P = \{00, 1101, 0001, 0111, 001, 1, 10,
1100, 111, 1010\}$ and $N = \{\epsilon, 0, 0000, 0011, 01, 010, 011,
100, 1000, 1001, 11, 1110\}$, the top row of Table
\ref{table_gpu_cpu_comparison}.  The table below shows the dependency
of synthesis cost (quantified by number of regular expressions checked
for compliance with the specification) on allowed error. We used the
(1, 1, 1, 1, 1) cost function. \\[0.5mm]

\begin{center}
{\small
 \input{data/table_allowed_error}
}
\end{center}
\vspace{2.5mm}

\NI What we see  might be an exponential dependency between allowed
error and synthesis cost. 

Finally, we challenge the deep learning community to solve minimal and
precise REI with ANNs, and then compare speed, memory usage and
power consumption with \ALGO.

%% file: data/table_allowed_error.tex
\begin{tabular}{|cl|c|c|c|}
	
	\toprule
	
  & Allowed Error  & \CARD{REs} & RE & Cost(RE)\\
  
	\midrule
	
	& 0 \% &   26,774,099,142 &  10?+0?(00+10*10?(0+1))1? &  28 \\
	& 5 \% &   319,649,322 &  ((0+1)0+(0+11)*1)(100?)? &  22 \\
	& 10 \% &   18,698,767 &  (10+0*1)(10?(0+1))? &  18 \\
	& 15 \% &   794,598 &  (0+1)0+(0+11)*1 &  14 \\
	& 20 \% &   116,912 &  (0+11)*(1+00) &  12 \\
	& 25 \% &   2,073 &  (0+11)*1 &  8 \\
	& 30 \% &   2,073 &  (0+11)*1 &  8 \\
	& 35 \% &   1,124 &  1+(0+1)0 &  7 \\
	& 40 \% &   50 &  10? &  4 \\
	& 45 \% &   3 &  1 &  1 \\
	& 50 \% &   1 &  $\emptyset$ &  1 \\

	\bottomrule
	
\end{tabular}

%% file: acknowledgement.tex
\section*{Acknowledgements}
We thank Richard Prideaux Evans for many discussions about
program synthesis, Akmal Ali, Hamish Todd, Emanuele Dalla Longa, John
Howson, and Marti Anglada for help with GPUs and GPU programming, and
Paul Kelly, Andrea Mondelli, Jim Whittaker for many discussions about
processor architecture in general, and the cost of data movement in
particular. Felix Dilke pointed us towards formal power series. We
thank the authors of
\cite{LeeM:synregefefiaa} for answering our questions about
AlphaRegex, and Daniel J\"unger for help with WarpCore.  We thank Amir
Naseredini for helping to prepare artefacts for submission, and Fardjad Davari
and Ali Momen Sani for answering software engineering questions.
Finally, we thank Mukund Raghothaman and the anonymous referees for providing
invaluable criticism, which greatly improved the presentation of this paper.

%% file: data-availability.tex
\section*{Code and Data Availability}
Our evaluated artefacts are not available in immutable archival format. All
benchmarks, source code and data are available from \cite{ArtifactGithub}.